\documentclass[journal]{IEEEtran}

\usepackage{placeins}
\usepackage{amsmath}
\usepackage{algorithm}
\usepackage{algorithmic}
\usepackage{booktabs}
\usepackage{multirow}
\usepackage[utf8]{inputenc} 
\usepackage[T1]{fontenc}    
\usepackage{hyperref}       
\usepackage{url}            
\usepackage{booktabs}       
\usepackage{amsfonts}       
\usepackage{nicefrac}       
\usepackage{microtype}      
\usepackage{xcolor}         
\usepackage{graphicx}
\usepackage{float}
\usepackage[numbers,sort&compress]{natbib}
\usepackage{caption} 

\newfloat{figtab}{htb}{fgtb}
\makeatletter
  \newcommand\figcaption{\def\@captype{figure}\caption}
  \newcommand\tabcaption{\def\@captype{table}\caption}
\makeatother

\def\BibTeX{{\rm B\kern-.05em{\sc i\kern-.025em b}\kern-.08em
    T\kern-.1667em\lower.7ex\hbox{E}\kern-.125emX}}

\title{EMD-Fuzzy: An Empirical Mode Decomposition Based Fuzzy Model for Cross-Stimulus Transfer Learning of SSVEP}
\label{sec:title}
\author{
\IEEEauthorblockN{
Beining Cao\textsuperscript{1}\textsuperscript{\dag}, 
Xiaowei Jiang\textsuperscript{1}\textsuperscript{\dag}, 
Daniel Leong\textsuperscript{1},
Charlie Li-Ting Tsai\textsuperscript{1},
Yu-Cheng Chang\textsuperscript{1}, 
Thomas Do\textsuperscript{1}, 
Chin-Teng Lin\textsuperscript{1}\textsuperscript{*}\\
}

\IEEEauthorblockA{\textsuperscript{1}GrapheneX-UTS Human-centric AI Centre, Australian AI Institute, School of Computer Science,\\ Faculty of Engineering and Information Technology, University of Technology Sydney}
\thanks{\textsuperscript{\dag}Beining Cao and Xiaowei Jiang contributed equally to this work.}
\thanks{\textsuperscript{*}Corresponding author: Chin-Teng Lin. Email: chin-teng.lin@uts.edu.au}
}

\begin{document}
\maketitle

\begin{abstract}

The Brain-Computer Interface (BCI) enables direct brain-to-device communication, with the Steady-State Visual Evoked Potential (SSVEP) paradigm favored for its stability and high accuracy across various fields. In SSVEP BCI systems, supervised learning models significantly enhance performance over unsupervised models, achieving higher accuracy in less time. However, prolonged data collection can cause user fatigue and even trigger photosensitive epilepsy, creating a negative user experience. Thus, reducing calibration time is crucial. To address this, Cross-Stimulus transfer learning (CSTL) can shorten calibration by utilizing only partial frequencies. Traditional CSTL methods, affected by time-domain impulse response variations, are suitable only for adjacent frequency transfers, limiting their general applicability. We introduce an Empirical Mode Decomposition (EMD) Based Fuzzy Model (EMD-Fuzzy), which employs EMD to extract crucial frequency information and achieves stimulus transfer in the frequency domain through Fast Fourier Transform (FFT) to mitigate time-domain differences. Combined with a Fuzzy Decoder that uses fuzzy logic for representation learning, our approach delivers promising preliminary results in offline tests and state-of-the-art performance. With only 4 frequencies, our method achieved an accuracy of \(82.75\% \pm 16.30\%\) and an information transfer rate (ITR) of \(186.56 \pm 52.09\) bits/min on the 40-target Benchmark dataset.
In online tests, our method demonstrates robust efficacy, achieving an averaged accuracy of 
\(86.30\% \pm 6.18\%\) across 7 subjects. This performance underscores the effectiveness of integrating EMD and fuzzy logic into EEG decoding for CSTL and highlights our method's potential in real-time applications where consistent and reliable decoding is crucial.

\end{abstract}


\section{Introduction}

\IEEEPARstart{B}{rain-computer} interface (BCI) is an innovative interaction technology that enables the direct, end-to-end transmission of user intentions from the brain to the controlled terminal \cite{fumanal2021motor, li2020sliding, lin2020direct}. Without the need for any physical involvement, BCI has garnered widespread attention due to its naturalness and intuitiveness \cite{samejima2021brain}. BCI can be broadly categorized into invasive and non-invasive types. And non-invasive BCI demonstrates greater potential due to its advantages of easy use and low risk \cite{edelman2024non}. Steady-state visual evoked potential (SSVEP) is an electroencephalography (EEG) signal elicited by visual stimulation with a specific frequency \cite{ke2023enhancing}. 
Brain signals with the same frequency can be induced when individuals stare at a flickering stimulus with a fixed frequency. And the intention of users can be detected by identifying which stimulus 
they are looking at. With its high accuracy and stability, SSVEP-based BCI has been applied in numerous real-world applications \cite{yin2014dynamically,guo2022ssvep,rivera2022cca}.

\begin{figure}[t]
    \centering
    \includegraphics[width=0.8\linewidth]{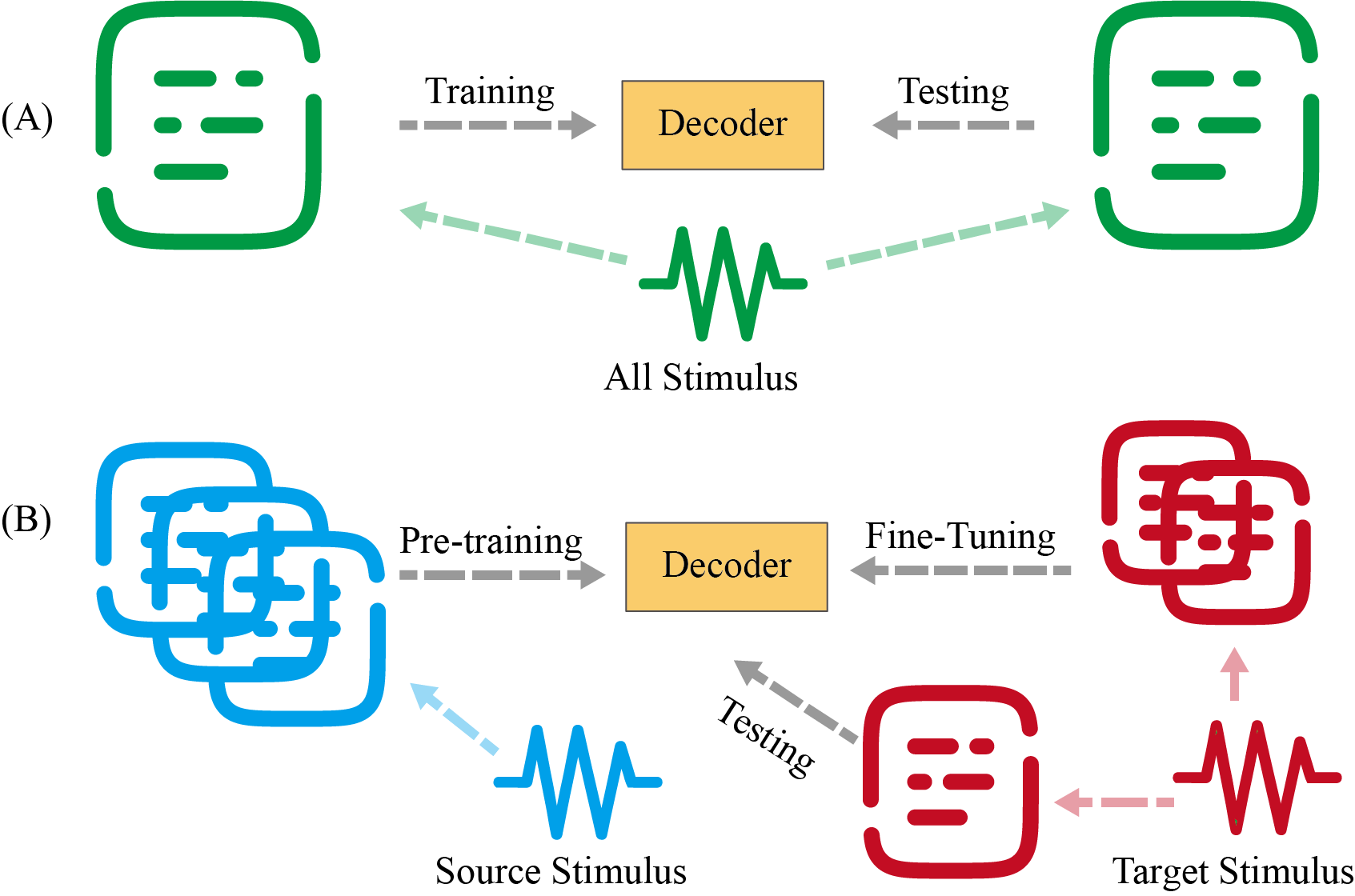}
    \caption{The diagram of three classification scenarios. 
    \textbf{(A): }intra-stimulus classification; 
    \textbf{(B): }inter-stimulus few-shot classification;
    }
    \label{fig:TLdemo}
\end{figure}

\begin{figure*}[htp]
    \centering
    \includegraphics[width=1\linewidth]{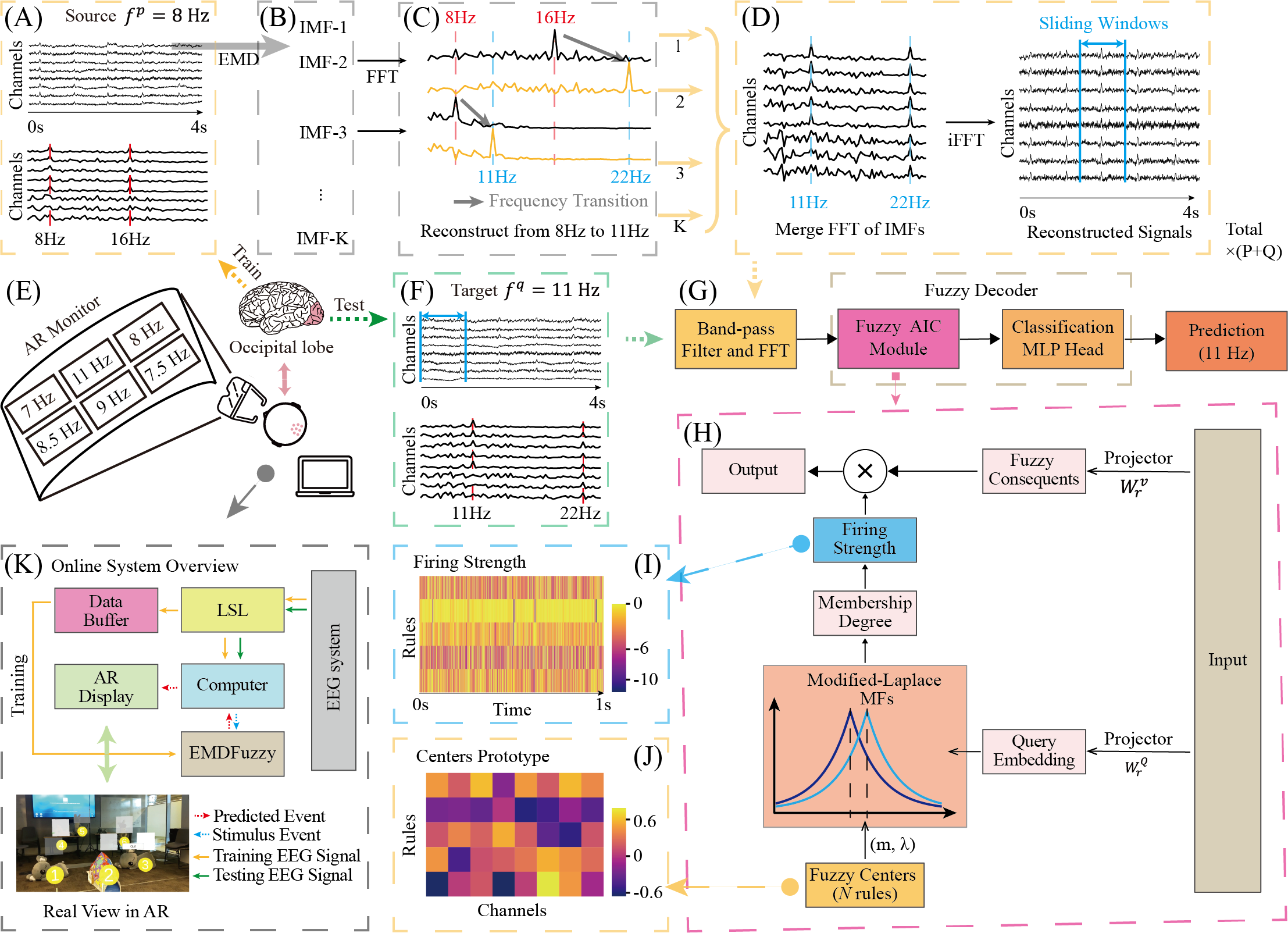}
    \caption{
    Overview of the online experimental setup and data processing in EMD-Fuzzy system. 
    \textbf{(A)} Training signal from an online experiment, showing a dominant 8 Hz source frequency and its 16 Hz harmonic in the spectrum. 
    \textbf{(B)} Decomposition of channel 1 into \(K\) intrinsic mode functions (IMFs) using empirical mode decomposition (EMD). 
    \textbf{(C)} Signal reconstruction in EMD-Fuzzy: Replacement of the target harmonic spectrum \(\Omega^p_{[1]}=22Hz\) with the source harmonic set \(\Gamma^q\) at 11 Hz, scaled by \(g^q\) in each IMF. 
    \textbf{(D)} Averaged reconstruction of IMFs post frequency-matching across channels, displayed alongside the reconstructed spectrum peaks for \(P=\{11Hz\}\) before iFFT transformation. After that, sliding windows (blue) are applied among all four seconds data.
    \textbf{(E)} Setup of the online experiment using Augmented Reality (AR) to display flickers at various frequencies; sensors over the occipital lobe are indicated in pink. 
    \textbf{(F)} Testing signal with target frequency 11 Hz and detection of its 22 Hz harmonic peak in the spectrum. Use only the first window (blue) as the test data.
    \textbf{(G)} Fuzzy Decoder architecture highlighting the Fuzzy AIC module and a classifier MLP head. 
    \textbf{(H)} Detailed structure of the Fuzzy AIC module: inputs are processed through Fuzzy Consequents and Query Embedding, with subsequent calculations of firing strength and membership degree via Fuzzy Center and Modified-Laplace membership functions (MFs), culminating in model output via TSK Fuzzy Model multiplication. 
    \textbf{(I)} Firing strength measurement from the fuzzy decoder for the testing signal. 
    \textbf{(J)} Learning of the Fuzzy Center from the reconstructed training signal. 
    \textbf{(K)} System design of the online setup, detailing EEG data acquisition, processing, and interface with the AR system and study participants during both training and testing phases; real AR view is shown in the lower sub-figure.
    }
    \label{fig:abstract_graph}
\end{figure*}

Currently, SSVEP decoding can be realized with either unsupervised or supervised learning techniques. Canonical cor-
relation analysis (CCA) \cite{lin2006frequency} and Filter-Bank CCA (FBCCA) \cite{chen2015filter} are two unsupervised methods for SSVEP. Although no training process is needed, these unsupervised decoders have poor performance when the EEG segment is short \cite{yuan2015enhancing}. 
Therefore, several supervised learning based decoders such as extended-CCA (ECCA) \cite{wong2020learning} and task-related component analysis (TRCA) \cite{nakanishi2017enhancing}, have been proposed, achieving high accuracy with shorter EEG segments. Compared with unsupervised methods, the supervised methods can get higher information transfer rate (ITR) \cite{obermaier2001information}. However, supervised learning methods require users to perform calibration experiments to collect EEG data elicited by each stimulus for model training, making the process highly time-consuming \cite{wong2020inter,bian2022small}. Particularly when there are numerous SSVEP targets, the lengthy calibration process can lead to user fatigue and even trigger photosensitive epilepsy symptoms \cite{cao2014objective}. Therefore, it is essential to reduce the calibration time to ensure a better user experience. 

A common SSVEP calibration strategy involves collecting EEG data from the user under each stimulation frequency for training, which is proved to be unnecessary recently. Based on the survey, existing studies \cite{wong2021transferring, wang2022stimulus} suggested that by collecting EEG data for only a subset of target frequencies and then reconstructing synthetic signals for the remaining frequencies through cross-stimulus transfer learning (CSTL), the model can be effectively trained as well. With CSTL, time cost of data collection can be reduced and the the calibration efficiency is enhanced. Currently, strategy for CSTL can be divided into three steps: signal decomposition, signal reconstruction, and model training. Firstly, the data of source domain (collected data) are decomposed in to several components which contain different information like frequency, phase or noise. Next, stimulus transfer is performed on components containing frequency information to generate components with the frequency of target stimulus. These components are then reassembled into signals of the target domain using signal reconstruction methods. Finally, the source domain and target domain data are combined for decoder training.

In details, Wong et.al proposed a transfer learning CCA (tlCCA) algorithm which uses 20 source stimulus to complete the model training for a 40-target SSVEP system \cite{wong2021transferring}. Based on the superposition theory, they leveraged a linear decomposition method to decompose SSVEP signal into two elements: the impulse response and the periodic impulse. For one individual, the neighboring stimulus frequencies share a common impulse response and different periodic impulse which determines the frequency and phase. Therefore, they combined the impulse response of source domain with the fabricated periodic impulses of target domain and use convolution to reconstruct synthetic signals for the target domain. Finally, they trained the spatial filters of decoders and get a accuracy over 80\% with 1s signals. However, data of 20 frequencies were leveraged because the author mentioned that tlCCA only work well when the frequencies of the source and target domain were close. The tlCCA method still requires a large amount of calibration data to ensure its accuracy. Wang et.al proposed a time-frequency-joint representation (TFJR) based CSTL method \cite{wang2022stimulus}. They developed a multi-channel adaptive decomposition Fourier decomposition with different phase (DPMAFD)  to decompose the source domain data into different components. Then the coefficients of common components across source domain were saved and the redundant noise components were eliminated. Finally, the inverse DPMAFD were used to reconstruct the target domain SSVEP by linearly combining the basic components of target stimulus with the common coefficients.  The TFJR achieved an accuracy of 80\% with only 8 source frequencies (1s EEG segment). However, it also leveraged the source signals closest to the target stimulus for reconstruction. When there is a significant difference between the source and target domain frequencies in SSVEP, the impulse response varies considerably in the time domain, making it difficult to reconstruct signals similar to the actual ones, thereby reducing the performance of CSTL. Therefore, it is crucial to address the issue of poor transfer performance caused by large frequency differences.

In addition, some other methods for signal decomposition and reconstruction have been applied in EEG field. EMD \cite{huang1998empirical} is a promising method for time-frequency analysis of EEG \cite{zheng2023enhancing}, enabling adaptive decomposition into the IMF components that reflect signal dynamics across different temporal scales. For SSVEP, EMD is effective for extracting IMF components containing clear frequency information \cite{chen2017new}. With the Fast Fourier Transform (FFT), the base and harmonic peaks corresponding to SSVEP signals can be observed in the spectrum of IMFs \cite{chang2022novel}. Transfer of stimulus can be achieved in the frequency domain based on the IMFs. Then the transferred IMF can be reconstructed into the target signal using signal recombination techniques like inverse FFT (iFFT) \cite{ouelha2017efficient}. 
The entire transfer process is achieved in the frequency domain, omitting the effects of signal differences in the time domain. Therefore, performing CSTL in the frequency domain may mitigate reconstruction difficulties caused by large intervals between source and target stimuli.  

Fuzzy Neural Networks (FNNs) have demonstrated their robustness in complex signal decoding tasks, such as EEG, due to their decision-making capabilities rooted in fuzzy logic systems. These systems can be trained using gradient descent optimization, making them an effective tool for signal processing \cite{lin1996neural, 10183374,106218}. In the context of transfer learning, fuzzy rule-based models offer the advantage of leveraging source domain knowledge through membership functions and rule centers, while retaining the knowledge in the form of human-understandable rules, which aids in decision-making. This is particularly beneficial for transfer learning across different domains or subjects, as demonstrated by iFuzzyTL \cite{jiangIFuzzyTLInterpretableFuzzy2024}, which effectively applies fuzzy logic to transfer learning in SSVEP task. The model draws on the principles of fuzzy logic \cite{luFuzzyMachineLearning2024}, specifically the Takagi–Sugeno–Kang (TSK) inference systems \cite{shihabudheen2018recent}.

In this paper, we developed a EMD-based CSTL module for data reconstruction of target stimulus and a fuzzy rule-based module as a decoder for SSVEP-based BCI systems. After the data reconstruction with EMD, the fuzzy model can effectively learn the mapping between the stimuli frequency and EEG signals, which is able to make accurate decisions by recognizing the match between the stimulus and the brain signal frequencies. Two open source SSVEP datasets were evaluated on the proposed method and several baseline methods. In addition, an augmented reality (AR)-based online test was conducted to prove the feasibility of the proposed method in real-world application.

The contributions of this study are outlined as follows:
\begin{enumerate}
    \item \textbf{Proposing an new data reconstruction method for CSTL:} With the EMD-based method, it is feasible and robust to generate effective training data for discriminative models using fewer collected data, thereby reducing the calibration costs of SSVEP systems.
    
    \item \textbf{Development of a novel fuzzy-based SSVEP decoder:} The proposed fuzzy model can effectively learn the frequency matching relationship between generated data and actual EEG, thereby enhancing the discriminative accuracy of the CSTL task.
    
    \item \textbf{Proving the applicability of the method:} An online experiment involving multiple subjects demonstrated the superior performance of our method, proving its feasibility in practical applications
\end{enumerate}

\section{Methods and materials}

\subsection{Explanation of SSVEP Principles and Stimulus Frequency Modulation}

In a SSVEP-based BCI system, when users are staring at a flickering stimulus with a fixed frequency \(f_j\) (\(j\)-th flicker), EEG signals with the base frequency \(f_j\) and Harmonic frequency $hf_j~(h \in H, \, h \in \mathbb{Z}^+$) can be synchronously elicited in their occipital lobes \cite{zhang2019fusing}. Users are supposed to select and focus on one stimulus, then their intention can be detected by identifying the frequency of the observed stimulus. 
For stimulation design, a variety of stimuli, such as flicker\cite{ming2023new}, Newton Ring \cite{xie2012steady} and reversal chessboard \cite{waytowich2016optimization}, have been proven to effectively elicit SSVEP. Among these stimuli, flicker paradigm is the most common used one due to its good performance. To build a flicker stimulus with a stable frequency, the sinusoidal encoding method is used to modulate the color of the flicker \cite{wang2016benchmark}, allowing its chromaticity to vary between black and white at a specified frequency:

\begin{align}
Ch(t) = \begin{bmatrix} 255 \\ 255 \\ 255 \end{bmatrix} \times \left( \frac{1 + \sin(2\pi f^j t + \theta^j_{Ch})}{2} \right)
\end{align}

where \(Ch(t)\) denotes the chrominance value at time \(t\), \(f^j\) is the label frequency, \(\theta^j_{Ch}\) is the phase shift, and \(j\) represents the number of target frequencies corresponding to \(j\)-th stimuli.

When users are looking at the \(j\)-th flicker, ideally, the elicited EEG \(x(t)\) in the visual cortex is:

\begin{align}
x(t,j) = \sum_{h=1}^{H^j} A^j_h \sin(2\pi h t + \theta^j_h) + \eta
\end{align}

where H and  \(h\)  are the set and element of harmonic waves for the elicited frequency \(f^j\) respectively, \(A^j_h\) and \(\theta^j_h\) denote amplitude and phase of \(h\) frequency, \(\eta\) represents the noise signals. Therefore, by decoding which frequency is generated, the intention of users can be recognized.

\subsection{Task Definition}

We consider a SSVEP task, where the goal is to perform transfer learning from a set of source stimuli \(P\) to target stimuli \(Q\). Let the set of all possible frequencies be denoted by \(J\), such that \( P \cup Q = J \) and \( P \cap Q = \emptyset \). The task involves leveraging data from the source stimuli \(P\) to train a model that generalizes well to the unseen target stimuli \(Q\).

Mathematically, we aim to build a function \(f: X_P \rightarrow Y_P\) using source frequency data \(X_P = \{x_p\}_{p \in P}\), where each \(x_p\) represents a feature vector from a source frequency \(p \in P\), and the corresponding label set \(Y_P\) contains the class labels for the source data. The goal is to transfer the learned representations to the target domain, \(X_Q = \{x_q\}_{q \in Q}\), where each \(x_q\) corresponds to an observation from a target frequency \(q \in Q\). 

In this setting, we employ a few-shot learning framework, where a small amount of data from the target frequencies is available during the training process. Specifically, we aim to minimize the loss function:
\[
\mathcal{L}(\varphi) = \sum_{p \in P} \mathcal{L}_P(f(x_p, \varphi), y_p) + \sum_{q \in Q} \mathcal{L}_Q(f(x_q, \varphi), y_q),
\]
where \(\mathcal{L}_P\) and \(\mathcal{L}_Q\) represent the loss functions for the source and target domains, respectively, and \(\varphi\) are the model parameters.

The objective is to fine-tune the model, initially trained on the source frequencies, to adapt to the target frequencies with limited data, thereby improving its performance on the unseen target frequencies through transfer learning. The challenge lies in effectively transferring the knowledge from \(P\) to \(Q\), given the limited availability of target data, and ensuring that the model generalizes well across both domains.

\subsection{Application of EMD for Efficient Analysis of SSVEP Signals}

EMD is highly effective for analyzing non-linear and non-stationary signals, such as EEG \cite{zhou2022empirical}. By decomposing a signal \( x(t) \) into IMFs and a residue, EMD provides a detailed frequency analysis of the signal. The decomposition can be expressed as follows \cite{ZHANG2021105572}:

\begin{equation}
    x(t) = m_k[x](t) + \sum_{k=1}^K d_k[x](t),
\end{equation}

where \( m_k[x](t) \) is the trend or mean of the signal, and \( d_k[x](t) \) represents the IMFs of the signal. The IMFs capture the different frequency components of the signal, allowing for a detailed analysis. 

The SSVEP signal \( x(t) \) can be ideally represented as a sum of sinusoidal components, each with time-varying amplitude and frequency, as follows:

Specifically, each IMF \( d_k[x](t) \) can be expressed as a sum of sinusoidal components, capturing the frequency content at various scales of the signal:

\begin{equation}
    d_k[x](t) = \sum_{s}^{S^j_k} A^j_s \sin(2\pi s t + \theta^j_s) + \eta_k
\end{equation}

where \( S^j_k \) is a subset of \( H^j \), representing all the frequencies present in the \( k \)-th IMF for the elicited frequency \(f^j\). $\eta_{k}$ is the noise component of the \( k \)-th IMF.This decomposition allows for a more detailed understanding and manipulation of the signal's frequency components.

In the context of SSVEP signals, this methodology is particularly useful. The SSVEP signal can be defined as a sum of sinusoidal components corresponding to different stimulus frequencies, and the IMFs obtained from EMD help to isolate and analyze these frequency components in a more efficient manner.

\subsubsection{Frequency Matching and Transfer Learning}
For effective CSTL, we employ a frequency exchange operation \( \Phi \) that aims to adaptively match and transfer frequency components between different signals. Given source harmonic frequency set \( \Gamma^q \) and target harmonic frequencies \( \Omega^p \), where each contains \( N_H \) harmonics of the source stimulus \( p \) and target stimulus \( q \), \( \Phi \) operates by exchanging these frequency components within the Fourier domain of the IMFs. The operation is defined as:

\begin{align}
    \Phi\left( d_k[x](t), \Gamma^p, \Omega^q \right) = \eta_k + \\
    &\sum_{i}^{N_H} g^q A_{\Gamma^p_{[i]}} \sin\left( 2\pi \Omega^q_{[i]} t + \theta_{\Omega^q_{[i]}} \right) + \nonumber \\
    &\sum_{i}^{N_H} g^p A_{\Gamma^p_{[i]}} \sin\left( 2\pi \Gamma^p_{[i]} t + \theta_{\Omega^p_{[i]}} \right)
\end{align}

where the terms represent the summation over distinct frequencies and phases, with \( S_k \) being the set of frequency components, \( \Gamma^p \) and \( \Omega^q \) being specific harmonic frequency sets of the source and target frequencies, respectively. \( \Gamma^p_{[i]} \) and \( \Omega^q_{[i]} \) are the \(i\)-th harmonic frequency in \( \Gamma^p \) and \( \Omega^q \). \( A_s \), \( f_s \), and \( \theta_s \) represent amplitude, frequency, and phase for each component. $g^p$ and $g^q$ are the gain scale for the amplitudes of the source domain and the target domain, respectively.

This transformation facilitates the reconstruction of signals that incorporate learned frequency characteristics from one domain into another. The reconstructed signal \( x'_{\Gamma^q, \Omega^p}(t) \) is then formulated by summing the transformed IMFs as:

\begin{align}
    x'_{\Gamma^q, \Omega^p}(t) = \sum_{k=1}^{K^{n}_1} \Phi(d_k[x](t), \Gamma^q, \Omega^p)
\end{align}

where \( K^{n}_1 \) represents the number of IMFs from the first to the \(n\)-th used in the reconstruction.

\subsubsection{Dataset Construction}

The training dataset \( X_{train} \) is composed of the reconstructed signals corresponding to various source and target stimulus pairs, represented as:

\begin{align}
    X_{train} = \{ x'_{\Gamma^q, \Omega^p}(t) \mid p \in P, q \in Q \}
\end{align}

This approach ensures a comprehensive training set that enables models to perform both cross-stimulus analysis and synthesis. The training data is processed through a set of transformations, including Fourier-based techniques, to extract frequency-domain features. The test set, however, retains the raw signals without any reconstruction. All data is then processed by applying the Fast Fourier Transform (FFT) to convert the signals into the frequency domain, where they can be analyzed for distinguishing characteristics relevant to the task at hand.

\subsection{The proposed Fuzzy Decoder}

\subsubsection{Fuzzy Inference System}

To decode the signal \(x(t)\), we propose a novel Fuzzy Inference System (FIS) based on FNNs~\cite{lin1996neural} and a Fuzzy Attention Layer~\cite{jiang2024fuzzybasedapproachpredicthuman, jiangIFuzzyTLInterpretableFuzzy2024}, which is trained using gradient descent optimization. The fuzzy decoder assigns a membership degree \(\mu(x) \in [0, 1]\), where \(0\) represents no membership and \(1\) represents full membership, to quantify the degree of membership of an element \(x\) in a fuzzy set \(\mu\).

In the Takagi-Sugeno-Kang (TSK) model, the relationship between the inputs and outputs is defined by a set of IF-THEN rules, as follows:

\begin{equation}
\text{If } x \text{ is } \mu_r, \text{ Then the output is } u_{r},
\end{equation}

where \(x\) is the input variable, \(\mu_r\) is the firing strength of fuzzy set \(r\), and \(u_r\) is the output associated with rule \(r\). The firing strength \(\mu_r\) is typically computed as the product of membership functions (MFs), which reflect the degree of membership of each feature in the rule. 

The final output of the TSK FIS is obtained through weighted normalization of the rule outputs:

\begin{align}
o = \sum_{j=1}^{R} \frac{\mu_j(x) u_j}{\sum_{i=1}^{R} \mu_i(x)},
\label{eq.rule_out}
\end{align}

where \(R\) is the total number of rules, and \(o\) represents the aggregated output.

\subsubsection{Fuzzy Modified-Laplace Membership Functions and Firing Strength}
To better capture the frequency peaks in SSVEP signals, we propose the use of Modified-Laplace Membership Functions (MFs), which offer advantages such as a sharper peak (higher sensitivity), increased robustness to outliers, and simpler gradient computation \cite{jiang2025iFuzzyAffduo}, making them more suitable for this task.

The Laplace distribution is defined as:
\begin{equation}
f(x \mid m, b) = \frac{1}{2b} e^{-\frac{|x - m|}{b}},
\end{equation}
where \( m \) is the location parameter and \( b \) is the scale parameter. To align with fuzzy membership function definitions and facilitate gradient computation, we modify the Laplace distribution by transforming the division into multiplication for the width parameter \( \lambda \):
\begin{equation}
\mu_{ML}(x \mid m, \lambda) = e^{-\lambda |x - m|},
\end{equation}
where \( \lambda \) controls the sensitivity of the MF and lies within the range \( \lambda \in [0, +\infty) \).

Within our Takagi-Sugeno-Kang (TSK) fuzzy system, the membership degree \( \mu_r(x) \) for rule \( r \) is computed as the product of Modified-Laplace MFs across features:
\begin{equation}
\mu_r(x) = \prod_{d} e^{-\lambda_d |x_{d} - m_{r,d}|},
\end{equation}
where \( x_d \) denotes the \( d \)-th feature of the input vector, and \( m_{r,d} \) is the center of the fuzzy set for the \( d \)-th feature under rule \( r \). The parameter \( \lambda_d \) dynamically adjusts the width of the MF for each feature.

The firing strength \( \overline{f_{ML}}_{i,r}(\mathbf{x}) \) for a fuzzy rule is calculated as the product of the membership degrees:
\begin{align}
\label{eq:fs}
\overline{f_{ML}}_{i,r}(\mathbf{x}) &= \frac{\mu_r(\mathbf{x})}{\sum_{j=1}^R \mu_j(\mathbf{x})} \\
&= \frac{e^{-\sum_{d=1}^D \lambda_{i,d} |x_{i,d} - m_{r,d}|}}{\sum_{j=1}^R e^{-\sum_{d=1}^D \lambda_{j,d} |x_{i,d} - m_{r,d}|}} \\
&= \text{softmax}_{i,r}\left(-\sum_{d=1}^D \lambda_{i,d} |x_{i,d} - m_{r,d}|\right),
\end{align}

\subsubsection{Fuzzy Attention Structure as Adaptive Linear Combiner System}

Consider an input \( x(t) \) processed by an adaptive linear combiner (ALC) system. The output \( Y(t) \) at time \( t \) is modeled as:

\begin{equation}
Y(t) = W_t^T \cdot x(t),
\end{equation}

where \( W_t \) represents the adaptive weights. These weights are derived from a TSK fuzzy model, with the firing strength \( \overline{f_r}(x(t)) \) corresponding to the input signal \( x(t) \). By incorporating the fuzzy attention mechanism, we assign the adaptive weights as \( W_t^T = \overline{f_r}(W_r^Q x(t)) \), where the \(W_r^V\) is the projection for the input signal \( x(t) \), and the output is then given by:

\begin{align}
        Y(t) = \overline{f_r}(W_r^Q x(t)) \cdot W_r^V x(t),
\label{eq:filter_design_time}
\end{align}

where the projection is represented by the parameter matrices \( W_r^V \) for rule \( r \). This formulation allows to adaptively modulate the importance of different features of \( x(t) \), based on their alignment with the fuzzy rule centers.

To ameliorate issues with gradient descent, we have modified the softmax operation in Equation (\ref{eq:fs}) for the firing strength to \texttt{log\_softmax}, which stabilizes the gradient flow by computing logarithms of probabilities. This adjustment output of our proposed fuzzy AIC module is formally expressed as:

\begin{equation}
Y_r(t) = \ln(\overline{f_{ML}}_{i,r}(W_r^Q x(t))) \cdot W_r^V x(t),
\end{equation}

\subsubsection{Fuzzy Decoder Structure}

The proposed fuzzy decoder consists of two primary components: the Fuzzy AIC module and a classifier head. For the Fuzzy AIC module, the input of our experiment is FFT features which retains the effective frequency-domain information, or raw EEG signals. The Fast Fourier Transform (FFT) of a time-series data \( \mathbf{g} \) computes both the real (Re) and imaginary (Im) components, which are essential for analyzing the frequency domain characteristics of the signal. The mathematical expressions for these components are as follows:

Given a time-series data vector \( \mathbf{g} \) with length \( L \), the FFT is defined by:
\begin{equation}
\text{FFT}(\mathbf{g}) = \sum_{k=0}^{L-1} \mathbf{d}[k] \cdot e^{-\frac{2\pi i}{L} k n}, \quad n=0,1,2,\dots,L-1
\end{equation}
where \( i \) is the imaginary unit. The real and imaginary parts of the FFT result for each frequency component \( n \) are extracted as follows:

\begin{align}
\text{Re}(n, ich) &= \operatorname{Re}\left(\text{FFT}(\mathbf{g})[n]\right) = \sum_{k=0}^{L-1} \mathbf{g}[k] \cdot \cos\left(\frac{2\pi k n}{L}\right) \\
\text{Im}(n, ich) &= \operatorname{Im}\left(\text{FFT}(\mathbf{g})[n]\right) = -\sum_{k=0}^{L-1} \mathbf{g}[k] \cdot \sin\left(\frac{2\pi k n}{L}\right)
\end{align}

The FFT features matrix \( X \) for $ ich$-th channel, which is used as the input for the Fuzzy AIC module, can be represented as follows:

\begin{equation}
X = \begin{bmatrix}
\text{Re}(0,0) & \cdots & \text{Re}(n,0) \\
\text{Im}(0,0) & \cdots & \text{Im}(n,0) \\
\text{Re}(0,1) & \cdots & \text{Re}(n,1) \\
\text{Im}(0,1) & \cdots & \text{Im}(n,1) \\
\vdots & \ddots & \vdots \\
\text{Re}(0,N_{ch}) & \cdots & \text{Re}(n,N_{ch}) \\
\text{Im}(0,N_{ch}) & \cdots & \text{Im}(n,N_{ch})
\end{bmatrix}
\end{equation}
where the $N_{ch}$ is the number of channels.

The classifier head, designed as a two-layer Multi-Layer Perceptron (MLP), processes the outputs of the fuzzy module for decision-making. The first layer of the MLP captures non-linear interactions among the fuzzy-processed features, followed by a ReLU activation function. The second layer then processes these activated features to output the predicted class probabilities.

\subsection{Main System Architecture}

Our proposed system, EMD-Fuzzy, follows a two-phase process comprising an EMD-based data reconstruction module and a fuzzy decoder. In the Phase I, the signal is preprocessed using EMD, which decomposes the signal into IMFs to capture various frequency components. The reconstructed signal, averaged by the \( K^{n}_1 \) IMFs, serves as the input for the second step.

The Phase II involves the fuzzy decoder, where the decomposed signal is passed through the Fuzzy AIC module. Here, fuzzy logic is used to assign membership degrees to each feature, dynamically adjusting the weights of each input component based on its relevance. These weighted features are then processed by the classifier head to generate the final prediction. 

This two-phase architecture ensures that both signal characteristics and fuzzy logic-based decision rules are effectively integrated for robust performance.

\subsection{Evaluation Metrics}

For evaluating the proposed method on SSVEP dataset, accuracy (ACC) and information transfer rate (ITR) are calculated as metrics. ACC is a fundamental measure that indicates the proportion of instances correctly classified out of the total instances, defined by the formula:
\begin{equation}
\text{ACC} = \frac{N_{correct}}{N_{trial}}
\end{equation}

where \(N_{correct}\) is the number of correctly classified trials, \(N_{trial}\) is the total number of trials.

On the other hand, ITR is a crucial metric for assessing the efficiency of classification systems, particularly in BCI systems. ITR considers both the speed and accuracy of classification to estimate the amount of information that can be transmitted per minute by the user. The formula for ITR is:

\begin{equation}
    \text{ITR} = \frac{60}{T_{total}} \left[ \log_2 N + P \log_2 P + (1 - P) \log_2 \frac{1 - P}{N - 1} \right],
\end{equation}

\begin{equation}
T_{total} = T_{staring} + T_{inference}
\end{equation}

Here, $T_{total}$ is the total time cost for one trial, which is composed of the time of user staring $T_{staring}$ and the model computation time $T_{inference}$.
$N$ is the number of SSVEP classes and $P$ is the accuracy.

\section{Experiments and Results}
\subsection{Dataset}
In this work, two open source SSVEP datasets, Benchmark~\cite{wang2016benchmark} and 12JFPM~\cite{nakanishi2015comparison}, were tested on the proposed method and several baseline methods evaluating the performance of them.  The details of two datasets are shown in Supplementary Section I-A.

\subsection {Data preprocessing}
 For the preprocessing of the two datasets, considering the prominent harmonic components below the fourth order in SSVEP, a Chebyshev bandpass filter \cite{lutovac2001filter} ranging from 7 to 70 Hz was initially used to eliminate low-frequency and high-frequency noise in the data. The minimum attenuation $gstop$ in the stopband was selected as 40 dB. In addition, the data were processed with a 50 Hz notch filter to eliminate the power-line noise \cite{chen2015filter}. Considering the attention shifts of the subjects at the start of each trial, the first 0.14 seconds of data from each dataset were discarded \cite{li2023precise}. For Dataset 1, data of the initial 0.64 seconds (cue phase + attention shift) and the final 0.5 seconds (rest phase) were discarded. For Dataset 2, we discarded the initial 0.14 seconds (attention shift) of data.

\subsection{Baseline methods}

To evaluate the proposed method, another five baseline methods were tested on the same two datasets. To validate the performance of different approaches on the CSTL task, the proposed method was compared with two state-of-the-art methods, TFJR and tlCCA. Furthermore, to evaluate the impact of classifiers on the proposed method, one commonly used SSVEP classifier, ECCA, was tested on data generated by EMD. Finally, FBCCA was tested to compare our method with unsupervised learning method. The details of baseline methods are shown as Supplementary Section I-B.

\subsection{Details of Model Evaluation}
\subsubsection{Strategy for model training}

\begin{figure}
    \centering
    \includegraphics[width=1\linewidth]{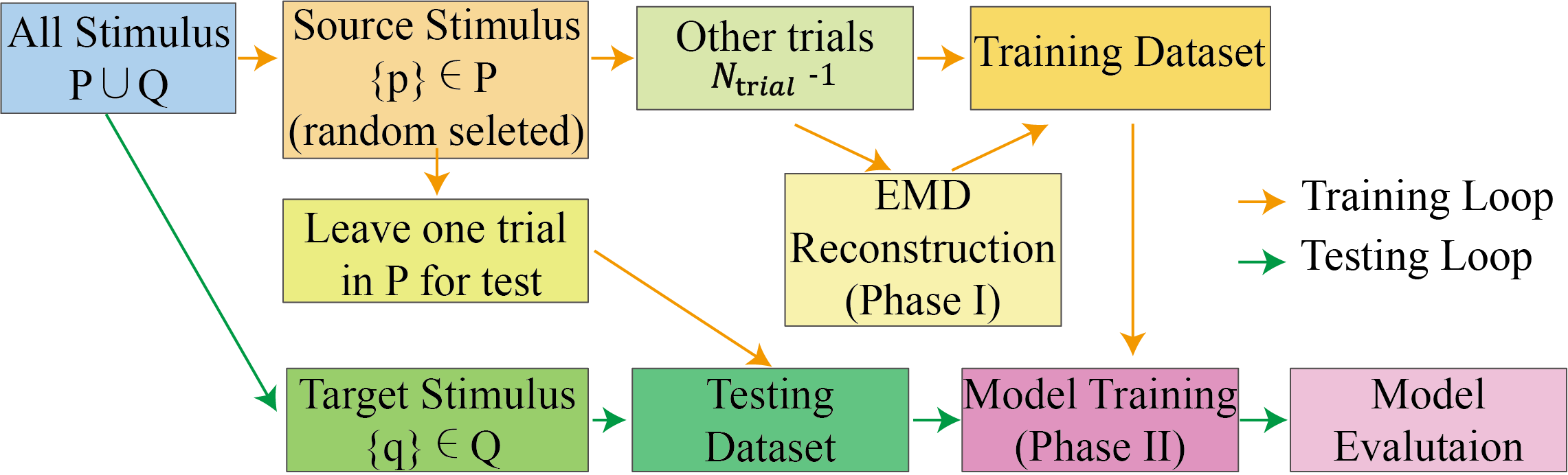}
    \caption{
    Schematic representation of the training and testing strategy for EMD-Fuzzy. All frequencies, \( P \cup Q \), are initially divided into source frequencies \( P \) (randomly selected) and target frequencies \( Q \) (remained). $N_{trial}$ denotes the number of trials for each class (frequency).
    }
    \label{fig:train_strategy}
\end{figure}

The details of our training strategy is as Fig \ref{fig:train_strategy}. Initially, from a total of $N_P + N_Q$ frequencies, $N_p$ frequencies are randomly selected as the source domain data, with the remaining $N_Q$ frequencies designated as the target domain. To evaluate the generalizability of model for recognition of all frequencies, one trial's data from the source domain dataset is randomly selected and combined with the target domain data to form the test set. This approach ensures that the test set contains SSVEP data corresponding to all frequencies. Then the remaining $N_{trial}-1$ trials from the source domain are used for data reconstruction.
During the data reconstruction process, we utilized each frequency ($P_p$) from the source domain to generate reconstructed data for all other frequencies (${Q_1},...,{Q_{N_Q}},{P_1},...{P_{p-1}},{P_{p+1}},...,{P_{N_P}}$). The reconstructed data is then merged with the remaining $N_{trial}-1$ trials from the source domain to serve as the training set for classifier training. Finally, the trained model is employed to classify and output results for the test set.  

In the CSTL task, the number of frequencies in the source domain might impact the results. To evaluate the effect of the number of source frequencies, for dataset 1, 4, 8, 12 and 20 frequencies from a total of 40 were randomly selected to build the source domain, with the remaining frequencies assigned to the target domain. For dataset 2, 2, 4, and 6 frequencies were randomly selected from a total of 12 to serve as the source domain.

\subsubsection{Implementation Details}
To eliminate the randomness of the results, thirty tests were repeatedly conducted for each model and used the average result as the final outcome. 

For the calculation of complex spectrum features, a frequency resolution of 0.23 Hz for 12JFPM was used, with the FFT range set between 6-64 Hz. Considering the impact of the number of rules in the Fuzzy model on discrimination accuracy, this study tested Fuzzy models with 3, 5, and 10 rules respectively. Given the limited size of the original dataset, a dynamic sliding window strategy was applied to the training set data to generate more training samples. 

The training framework operates on a structure of epoch-based iterations, with an upper limit of 100 epochs. The Adamw optimizer \cite{loshchilov2018decoupled}, known for effectively managing sparse gradients and including a weight decay coefficient of 0.01 for enhanced regularization, is employed. It is configured with adaptive learning rates, with beta coefficients set at 0.9 and 0.95, controlling the exponential decay rates of the moving averages. Additionally, an epsilon value of $1 \times 10^{-8}$ is incorporated to ensure numerical stability during computations.

\subsection{Model Comparisons}

\begin{figure*}
    \centering
    \includegraphics[width=1\linewidth]{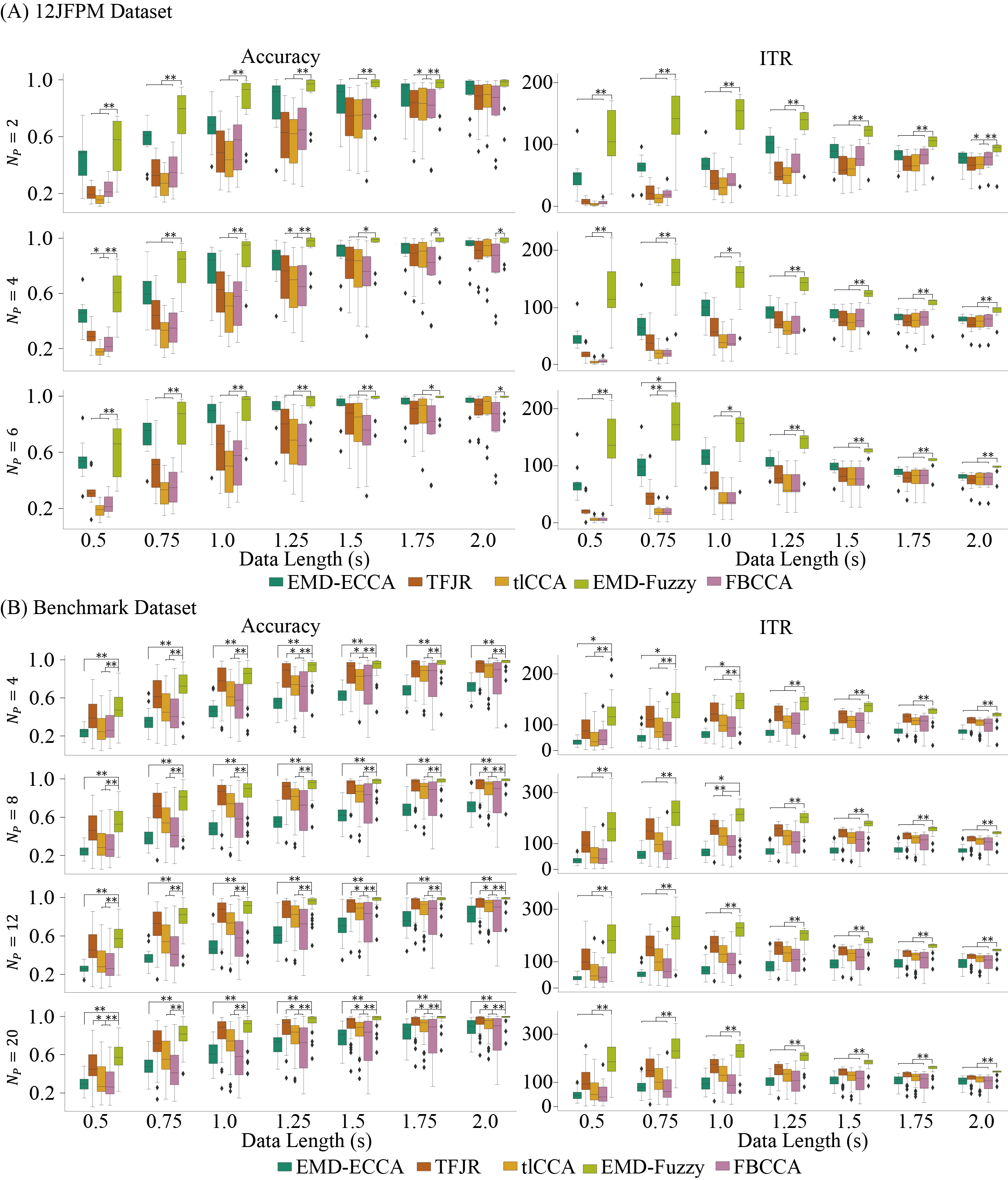}
    \caption{Results of model comparisons across various data lengths (from 0.5s to 2s, in increments of 0.25s). \textbf{(A):} Accuracy and ITR of the 12JFPM Dataset. \textbf{(B):} Accuracy and ITR of the Benchmark Dataset. Statistical significance between our proposed EMD-Fuzzy model and baseline models is indicated by * ($p<0.05$) and ** ($p<0.01$).}
    \label{fig:res_model_comparisons}
\end{figure*}

We investigate the influence of the number of query points $N_P$ on model performance across two datasets. Specifically, we vary $N_P$ in the 12JFPM dataset at levels 2, 4, and 6, and in the Benchmark dataset at levels 4, 8, 12, and 20. Each configuration randomly splits the sets $PQ$ and $Q$ and evaluates across different data lengths.

For the 12JFPM dataset, the EMD-Fuzzy model demonstrated consistent improvement as data length increased. At a data length of 0.5 seconds, the model achieved an average accuracy of \(53.42\% \pm 20.27\%\), escalating to \(72.43\% \pm 22.58\%\) at 0.75 seconds, \(83.42\% \pm 21.34\%\) at 1 second, \(89.87\% \pm 16.28\%\) at 1.25 seconds, and \(92.11\% \pm 14.14\%\) at 1.5 seconds with $N_Q=2$. These improvements are statistically significant ($p<0.05$), affirming state-of-the-art (SOTA) performance as depicted in Fig. \ref{fig:res_model_comparisons}(A).

In the Benchmark dataset, similar trends were observed as data lengths increased. Starting at \(49.22\% \pm 17.06\%\) for 0.5 seconds, accuracy improved to \(70.82\% \pm 18.14\%\) at 0.75 seconds, \(80.61\% \pm 16.76\%\) at 1 second, \(90.05\% \pm 12.48\%\) at 1.25 seconds, \(92.96\% \pm 10.89\%\) at 1.5 seconds, \(94.75\% \pm 10.54\%\) at 1.75 seconds, and culminated at \(95.69\% \pm 12.02\%\) at 2 seconds with $N_Q=4$. These results also highlight SOTA performance with significant enhancements ($p<0.05$) as shown in Fig. \ref{fig:res_model_comparisons}(B).

\section{Real-Time Feasibility Evaluation}

\begin{figure}[ht]
    \centering
    \includegraphics[width=\linewidth]{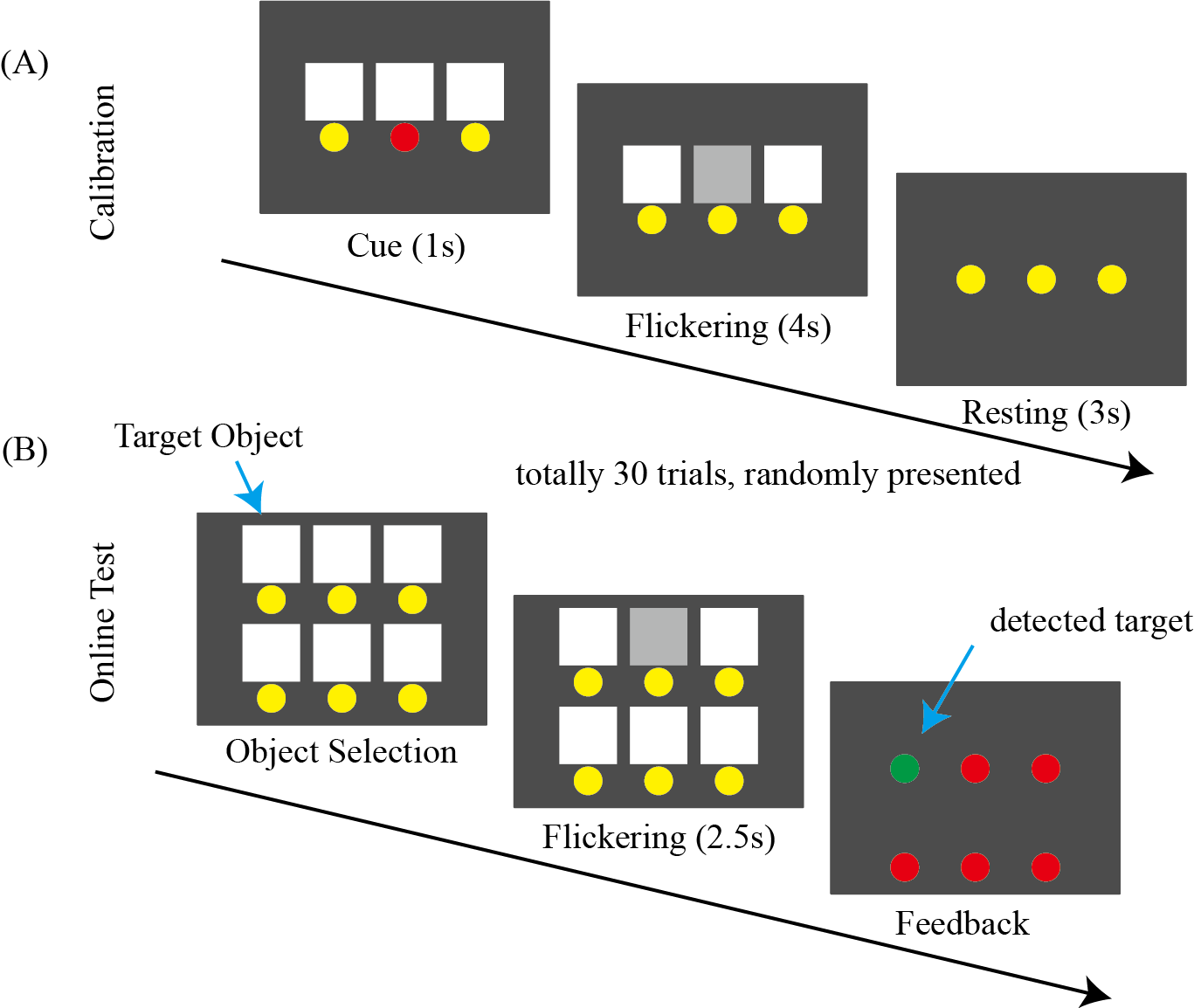}
    \caption{Schematic of the real-time experiment design, illustrating the setup for the calibration phase (A) and the real-time testing phase (B). This layout details the sequence of tasks, participant interactions, and data collection points critical for evaluating the real-time performance of the system.}
    \label{fig:online_exp_design}
\end{figure}

To verify the practicality of the method proposed in this study for real-time systems, an AR-based online SSVEP experiment was designed. In this experiment, participants were required to select objects by staring at the SSVEP flickers on the objects. In this real-time system, six flickers with frequencies including 7Hz, 7.5Hz, 8Hz, 8.5Hz, 9Hz and 11Hz were set for object selection.

\subsection{System Implement}
The composition of the real-time SSVEP system is illustrated as Fig. \ref{fig:abstract_graph} (E) and (K). In this experiment, HoloLens2 headset (Microsoft Corporation) was used to build an AR environment for displaying the SSVEP stimuli. The refresh rate of HoloLens2 was selected as 120 Hz, which was high enough to maintain the stable flickering. A 8-channel MentaLab dry sensors (Mentalab Technology LLC) was used for EEG collection. The sampling rate was selected as 500 Hz. A button electrode was attached to the mastoid area, which served as reference electrodes connected to the sensor. A Dell latitude 5540 computer (13th Gen Intel(R) Core(TM) i7-1370P 1.90 GHz) was used for data acquisition and processing. 

In the entire system, EEG data were acquired by the sensor and transmitted to the computer via the Lab Streaming Layer (LSL) protocol \cite{kothe2014lab}. The computer then saved the data for the calibration phase and invoked the model for classification in the online-test phase. The online results were sent to the HoloLens2 through User Datagram Protocol (UDP) \cite{postel1980rfc0768} to be displayed to the participants as feedback.

\subsection{Experiment details} \subsubsection{Procedure of calibration}
In calibration phase, a few trials of data  were collected for EEG reconstruction and model training. In the calibration phase, three frequencies including 7Hz, 8Hz and 9Hz were selected as the source domain. Data of these three frequencies were collected and used for reconstructing the SSVEP of 7.5Hz, 8.5Hz and 11Hz. 

The calibration procedure is illustrated as Figure \ref{fig:online_exp_design} (A).
In the initialization interface of the experiment program, users can position the three flickering blocks anywhere within their field of vision by virtually dragging them, ensuring that each flicker can be clearly seen. Followed by this, subjects were required to press a virtue "Start" button in AR to activate the system. Subsequently, the circle beneath one of the flickers changed from yellow to red and lasted for 1 second, indicating that the subject should look at that flicker during this trial. Then all three flickers began to flash and the subject stared at the cued flicker for 4 seconds. Finally, a 3 seconds rest was given to subjects for adjusting their status. In each trial, a flicker was randomly selected as the stimulus. Each subject underwent thirty calibration trials, with ten trials for each category of flicker.

\subsubsection{Online Test}

Six flickers were used in online test for object selection. The procedure of online test is illustrated as Figure \ref{fig:online_exp_design} (B). Initially, subjects were required to put six flickers freely on the object in the real-world scenario. Then they pressed the "start' button and began the online test. Then all flickers started to flash, and subjects stared at the one on the object they wish to select. To ensure a better user experience, the duration of the flicker in the online experiment has been reduced to 2.5 seconds. Finally, the detected result was sent to AR and displayed as a green circle for 3 seconds. 
subjects said "Yes" to inform the experimenter when they got correct feedback; otherwise, they said "No" to indicate an incorrect recognition. The experimenter recorded the results of all trials. The online experiment comprised 24 trials across two sessions. Each participant was given a 5-minute rest period after completing the first 12 trials and finished the remaining 12 trials in the second session. Within the 24 trials, the flicker for each frequency was selected for four times.

\subsubsection{Subject and environment}

Five subjects including 4 males and 3 females participated the online test. They ranged in age from 26 to 31 (\(27.714~\text{year} \pm 2.05\)). Two of them had experience in SSVEP experiment and the other subjects were naive. All participants were in good mental health and had normal or corrected-to-normal vision. Prior to the commencement of the experiment, each subject signed the ethical consent file that was approved by the University of Technology Sydney's Ethical Committee (Grant number: UTS HREC REF No. ETH20-5371). 

The experiment was conducted under the background of a bright conference room setting, where six items including dolls and model sculptures were placed at various locations as objects. Participants were seated in an adjustable chair, positioning themselves at a height that allowed full visibility of all objects within the AR environment. The distances between the subjects and the six items varied from 1.5 meters to 3 meters. In order to eliminate the impact of object position on the participants, the positions of the objects were maintained consistently for all subjects. 

\subsection{Results of the Real-Time Test}

\begin{figure}
    \centering
    \includegraphics[width=1\linewidth]{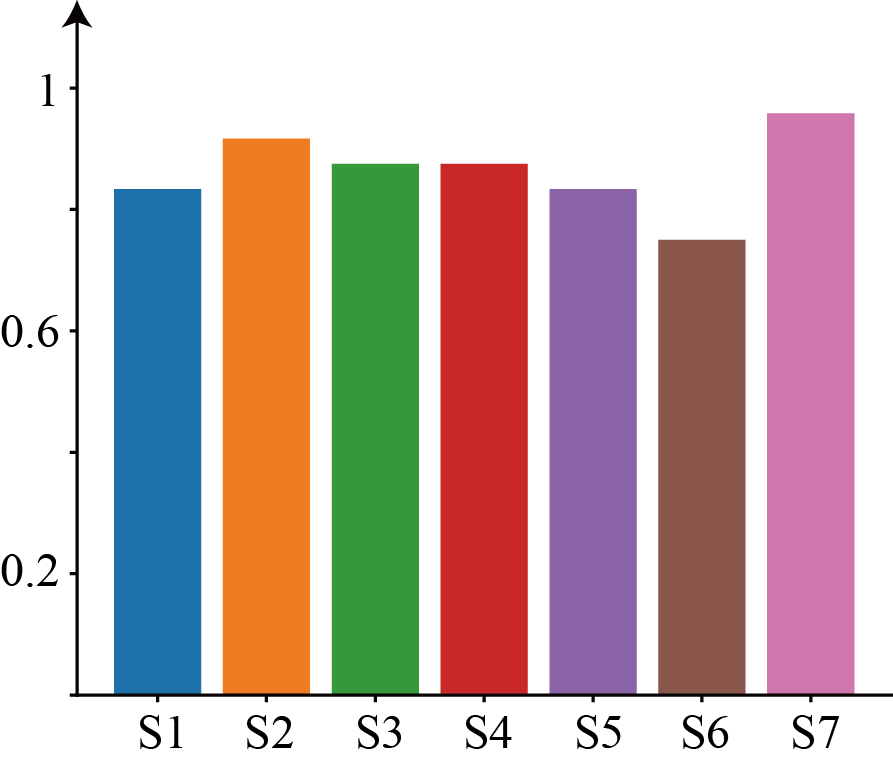}
    \caption{Bar graph illustrating the accuracy results from the 2.5-second real-time experiment.}
    \label{fig:onlineres}
\end{figure}

The accuracy of the 2.5-second real-time experiment was reported as \(86.30\% \pm 6.18\%\), with details shown in Fig. \ref{fig:onlineres}. This result indicates a robust performance under operational conditions, with a relatively small standard deviation, suggesting consistent responses across different subjects. 

The calibration time for one subject lasted for around 4 minutes, which included a data collection duration of 210 seconds and some preparation time. All subjects reported no adverse reactions such as fatigue or dizziness during the experiment. Therefore, this online experiment demonstrates that our method can enhance the user experience by reducing calibration time.

\section{Discussion}

\subsection{Spectral Similarity Analysis Between Reconstructed and Original Signals}

The efficacy of the CSTL approach is contingent upon the degree of similarity between the reconstructed and original signal patterns. Given the reliance on frequency domain reconstruction for signals in the target domain, this study provides a detailed analysis of the spectral similarity between the reconstructed and actual signals. Here, a CSTL case using 8 source domain signals to reconstruct 32 target domain signals is selected for analysis.

\begin{figure}[htbp]
    \centering
    \includegraphics[width=0.5\textwidth]{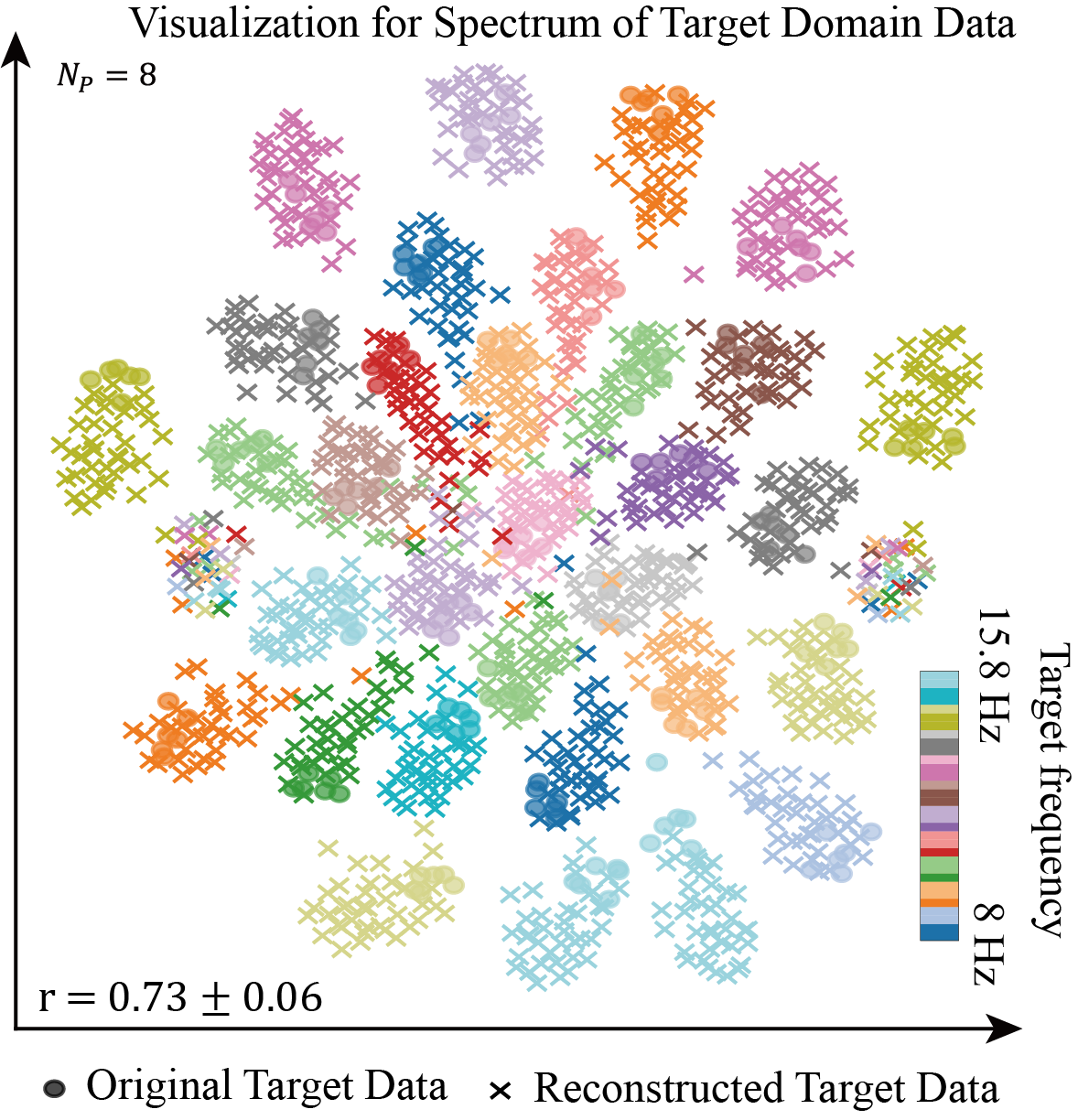}
    \caption{UMAP visualization of the spectral similarity between the original and EMD-reconstructed data. The close clustering of points indicates a high degree of correspondence.}
    \label{fig:emd_or_compare}
\end{figure}

Initially, the Pearson Correlation Coefficient (PCC) \cite{cohen2009pearson} is computed between the spectrum of real target data and their EMD reconstructions, yielding a value of $0.73 \pm 0.06$. This result indicates a high degree of similarity between the EMD-reconstructed spectrum and the real signal distribution. Figure \ref{fig:emd_or_compare} presents a UMAP visualization that further elucidates this similarity. Therefore, the high similarity between the reconstructed data and the real data enables the Fuzzy model to accurately learn the common patterns between them, thereby achieving precise discrimination.

\subsection{Demonstration of SSVEP Signal Inference by the Fuzzy Decoder}
In this section, we demonstrate how the Fuzzy Decoder infers the SSVEP signal. During the online test, signals in the test set have a frequency of \(f^q=11\,Hz\), as depicted in Fig. \ref{fig:abstract_graph}(F). The firing strength and center \(m\) are shown in Figs. \ref{fig:abstract_graph}(I) and \ref{fig:abstract_graph}(J), respectively. Notably, rule \#2 contributes most significantly in this case. 

\begin{figure}[htp]
    \centering
    \includegraphics[width=1\linewidth]{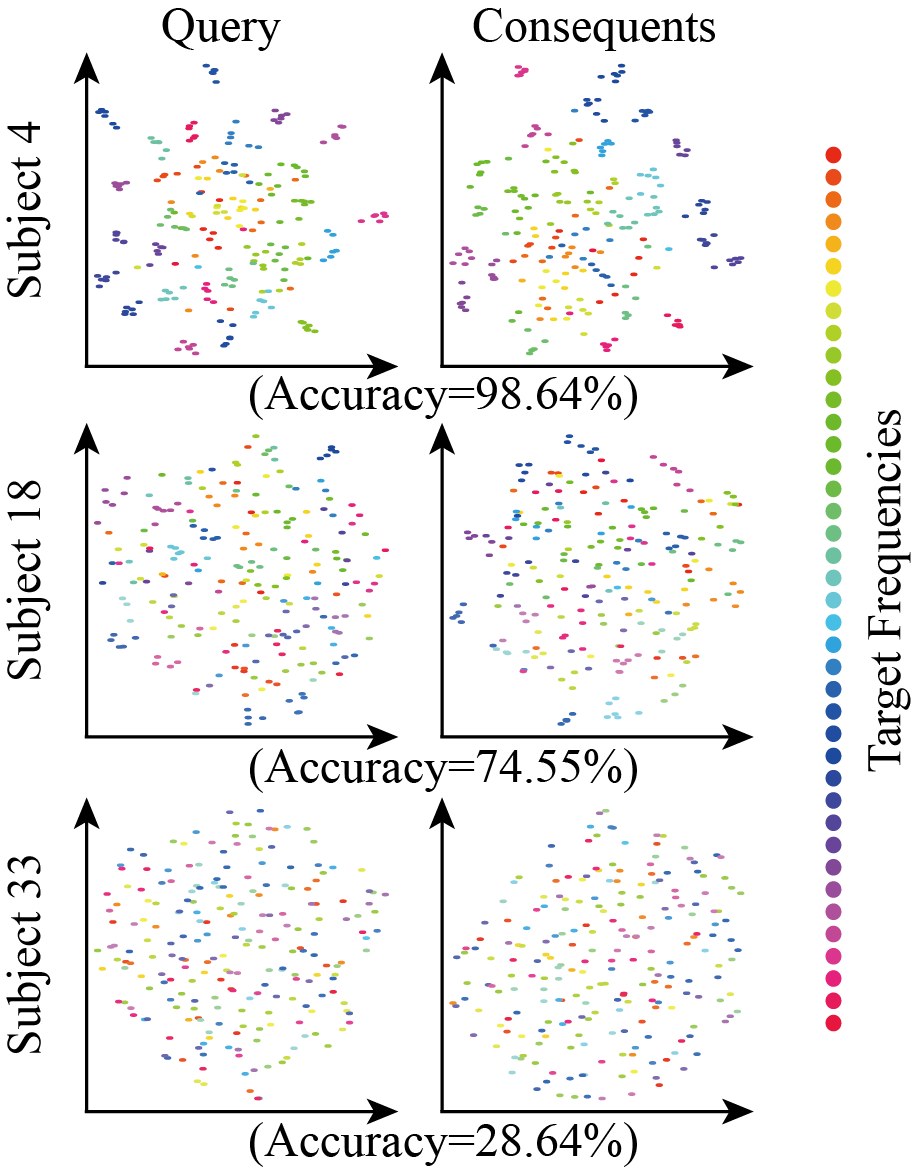}
    \caption{Visualization of UMAP embeddings for three subjects from the Benchmark dataset, illustrating the performance variability of our proposed EMD-Fuzzy's Query and Fuzzy Consequents Projectors.}
    \label{fig:umap_i}
\end{figure}

Additionally, we illustrate the learned embeddings from EMD-Fuzzy's Query and Fuzzy Consequents Projectors, which vary in performance levels, as shown in Fig. \ref{fig:umap_i}. The embeddings demonstrate the spatial distribution of target frequencies for each subject along with their corresponding classification accuracies: 98.64\% for Subject 4, 74.55\% for Subject 18, and 28.64\% for Subject 33. This visualization highlights how the quality of embeddings impacts model performance.

\subsection{Analysis of Error Cases in Fuzzy Decoder SSVEP Signal Interpretation}
In this section, we explore error cases observed during the online testing of our Fuzzy Decoder model. These error cases provide crucial insights into the limitations and potential areas of improvement for our model.

\begin{figure*}[htp]
    \centering
    \includegraphics[width=1\linewidth]{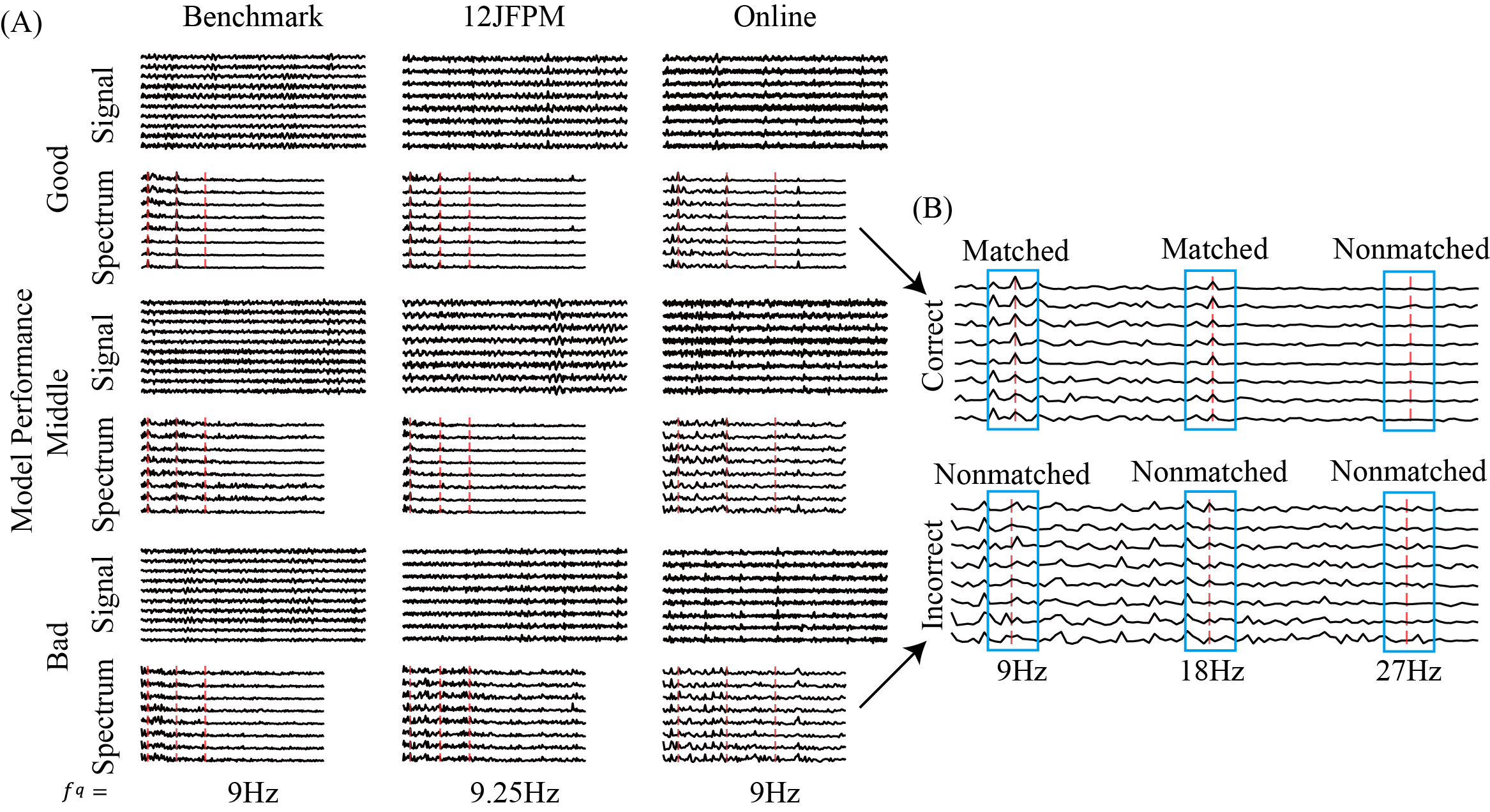}
    \caption{Case Analysis of Signal Performance. 
    \textbf{(A):} Demonstrations of three different performance levels across datasets. Each subfigure displays both the signal and its corresponding spectrum.
    \textbf{(B):} Comparative analysis of good and bad performance within our Online Dataset, illustrating reasons for model failures. In this erroneous case with \(f^q = 9\,Hz\), no prominent peaks are observed at the fundamental or harmonic frequencies (18 and 27 Hz), unlike the correct instances shown where the first harmonic frequency at 18 Hz is clearly visible.
    \label{fig:case-analysis}}
\end{figure*}

Fig. \ref{fig:case-analysis} exemplifies how different setups in the Fuzzy Decoder can influence the accuracy and effectiveness of the signal interpretation. Panel (A) of the figure shows three different performance levels—good, middle, and bad—across datasets, highlighting the model's ability to process and decode SSVEP signals under varying conditions. The corresponding spectra in each subfigure demonstrate how well the signal frequencies are captured by the model, with clear distinctions in signal clarity and spectral peaks between the performance categories.

Panel (B) offers a closer look at specific instances of good and poor model performance, providing an explicit comparison that illustrates why the model fails in certain situations. The example with \(f^q = 9\,Hz\) is particularly informative, as it shows a lack of prominent peaks at both the base frequency and its harmonics in the poor performance scenario, which contrasts sharply with the good performance scenario where the first harmonic frequency at 18 Hz is clearly discernible.

This analysis underscores the importance of robust feature extraction and the ability of the Fuzzy Decoder to accurately identify frequency components critical for correct SSVEP signal decoding. Understanding these error cases enables us to refine our approach, enhancing model reliability and accuracy in interpreting neural signals.

\subsection{Ablation of the Number of Source Frequencies}

In this section, we investigate the effects of varying the number of source frequencies on the performance of different algorithms in SSVEP-based BCI systems. The number of source frequencies, denoted as \(N_p\), is a critical factor influencing the performance of the system.

\begin{figure}
    \centering
    \includegraphics[width=1\linewidth]{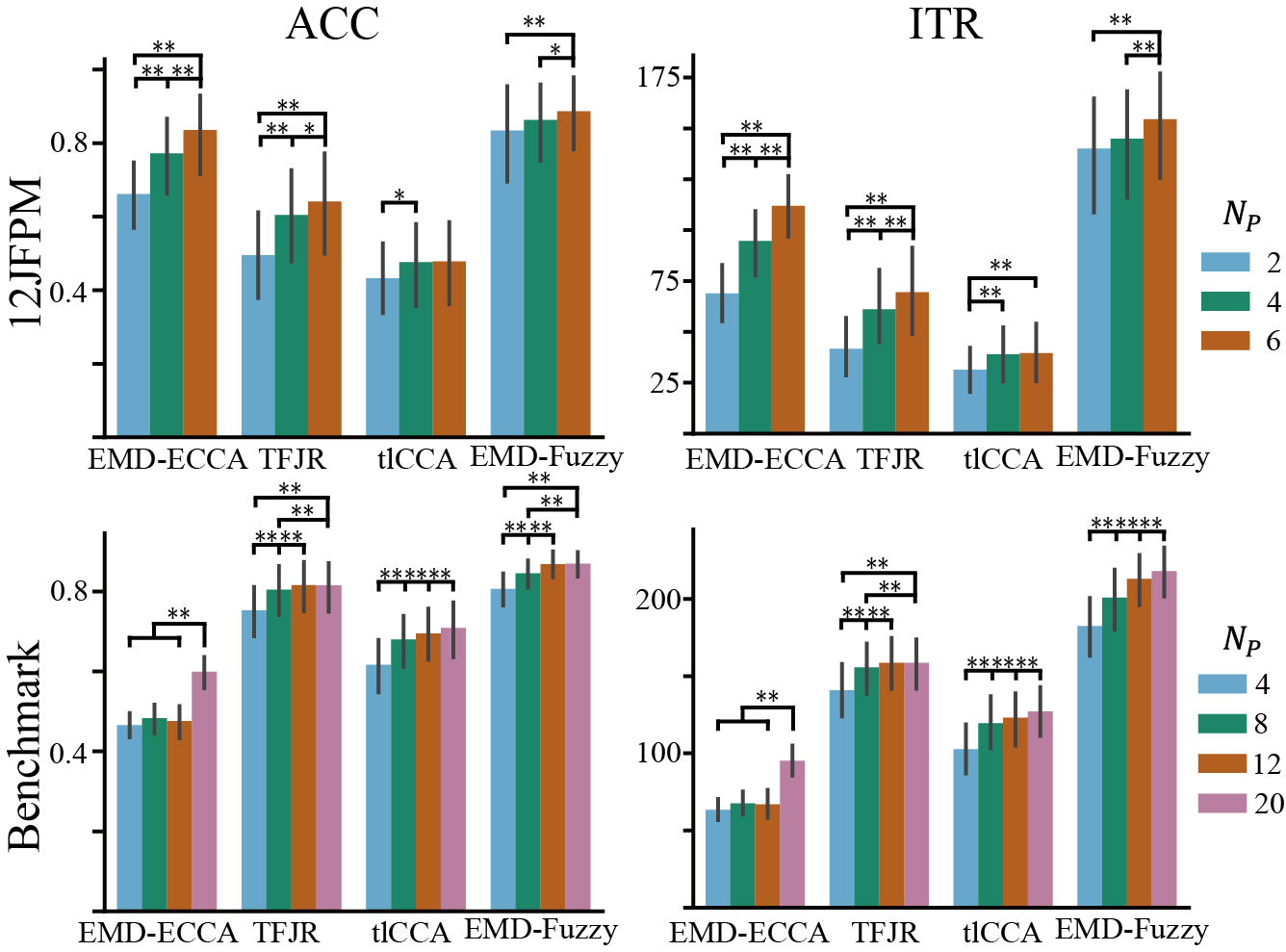}
    \caption{Performance comparison of different decoding techniques over a range of source frequencies $N_p$ for ACC and ITR. The left panel displays ACC and the right panel displays ITR for methods including EMD-ECCA, TFJR, t1CCA, and EMD-Fuzzy. (**p<0.01, ***p<0.001, after FDR-BH).}
    \label{fig:performance-comparison}
\end{figure}

As shown in Figure \ref{fig:performance-comparison}, both ACC and ITR metrics demonstrate increased performance with a rise in \(N_p\), indicating that the algorithms effectively utilize additional frequency information. Notably, ITR exhibits greater sensitivity to changes in \(N_p\), with significant performance increments observed across the datasets. The EMD-ECCA model shows less sensitivity to variations in \(N_p\) compared to our proposed EMD-Fuzzy model. This disparity may be attributed to the Fuzzy Decoder's superior ability to adaptively learn patterns using fuzzy logic-based center, which effectively mitigates the loss of information in true frequency, a challenge that more severely impacts ECCA's ability to extract useful signals.

Both TFJR and t1CCA display patterns similar to those of the EMD-Fuzzy model, indicating a comparable level of robustness in handling increased complexity in source frequencies. However, the EMD-Fuzzy model consistently outperforms the other methods, suggesting that its integration of fuzzy logic provides a distinct advantage in optimizing the decoding process.

\subsection{Ablation of the Rule Count in Fuzzy Decoder}
\begin{figure}
    \centering
    \includegraphics[width=1\linewidth]{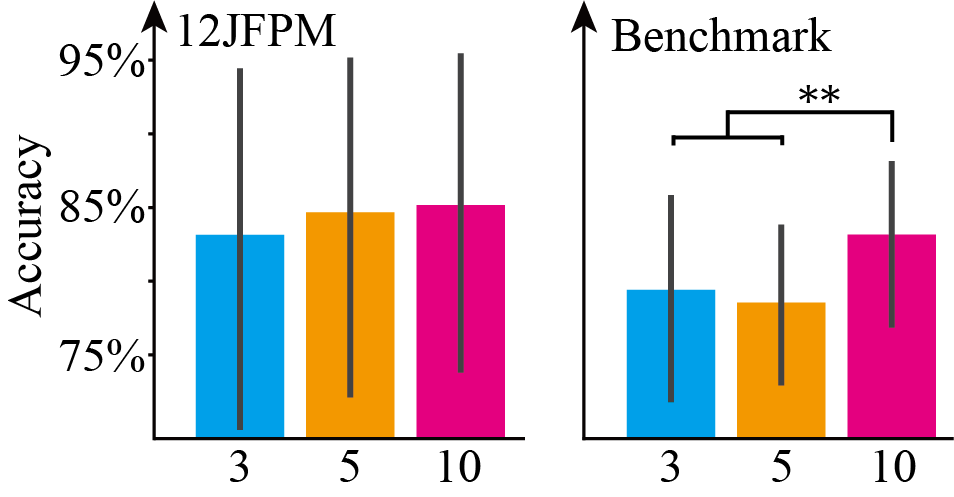}
    \caption{Performance comparison of fuzzy decoders with different rule counts (R=3, R=5, and R=10). The statistical significance is marked where applicable (**\textit{p}<0.01).}

    \label{fig:ab_nRule}
\end{figure}

To investigate the influence of rule count on the performance of our fuzzy decoder, we conducted an ablation study varying the number of rules (R = 3, 5, and 10). The statistical comparisons, detailed in Fig. \ref{fig:ab_nRule}, indicate a non-significant impact of increasing rule counts from 3 to 5 ($t = 1.13, \textit{p} = 0.29$) and 3 to 10 ($t = -1.29, \textit{p} = 0.23$) in the 12JFPM dataset, with performance metrics of $0.83 \pm 0.19$ and $0.85 \pm 0.16$, respectively. Similarly, changing from 5 to 10 rules did not result in significant differences ($t = 0.50, \textit{p} = 0.63$). However, in the Benchmark dataset, significant improvements were observed when increasing the rule count from 5 to 10 ($t = 5.58, \textit{p} < 0.01$) and from 3 to 10 ($t = 3.32, \textit{p} < 0.01$), suggesting that while the rule count can influence performance, the effects are not consistently substantial across different configurations.

\subsection{Limitations}
Although the CSTL method proposed in this paper achieves better results compared to existing studies, there are still some limitations. 
Firstly, the method proposed still exhibits significant dependency on the source domain data. When the number of frequencies in the source domain is limited, the accuracy of the transfer is not yet satisfactory. Therefore, it is necessary to incorporate some few-shot learning techniques based on this study to achieve better discrimination accuracy under the condition of limited source domain data. Secondly, the online experiments in this work utilized an SSVEP paradigm with six frequencies, which is relatively few for practical applications. Therefore, future work could introduce SSVEP commands with more frequencies to enhance the practical value of the proposed method. Third, current studies on CSTL in the SSVEP field are still based on individual-specific evaluation. The cross-subjects transferability of models has not yet been evaluated. Therefore, a cross-subject CSTL method is necessary to achieve a plug-and-play BCI.

\section{Conclusion}
This work proposes an EMD-Fuzzy method for the CSTL task in SSVEP applications. In this method, the source domain signals are decomposed into IMF components containing effective frequency information using EMD. Stimulus transfer are then performed in the frequency domain using FFT and iFFT, overcoming adverse effects caused by time-domain differences. A Fuzzy model is employed to learn the common frequency pattern information between the reconstructed signals and the actual EEG signals, achieving accurate discrimination of SSVEP signals.

The work was compared with baseline methods on two public datasets, achieving state-of-the-art results. Additionally, an online SSVEP test based on an AR environment was conducted, achieving an average accuracy of 86.30\% across seven subjects, validating the practicality of this method. The proposed approach requires only a subset of labeled EEG signals for training, significantly reducing calibration time for users and greatly optimizing the user experience for SSVEP-BCI systems.

\textbf{Acknowledgments:} This work was supported in part by the Australian Research Council (ARC) under discovery grant DP220100803 and DP250103612 and ITRH grant IH240100016, Australian National Health and Medical Research Council (NHMRC) Ideas Grant APP2021183, and the UTS Human-Centric AI Centre funding sponsored by GrapheneX (2023-2031). Research was also sponsored in part by the Australia Defence Innovation Hub under Contract No. P18-650825 and the Australian Defence Science Technology Group (DSTG) under Agreement No: 12549.

\bibliographystyle{IEEEtran}
\bibliography{ref, ref_Beining}

\begin{thebibliography}{10}
\providecommand{\url}[1]{#1}
\csname url@samestyle\endcsname
\providecommand{\newblock}{\relax}
\providecommand{\bibinfo}[2]{#2}
\providecommand{\BIBentrySTDinterwordspacing}{\spaceskip=0pt\relax}
\providecommand{\BIBentryALTinterwordstretchfactor}{4}
\providecommand{\BIBentryALTinterwordspacing}{\spaceskip=\fontdimen2\font plus
\BIBentryALTinterwordstretchfactor\fontdimen3\font minus \fontdimen4\font\relax}
\providecommand{\BIBforeignlanguage}[2]{{%
\expandafter\ifx\csname l@#1\endcsname\relax
\typeout{** WARNING: IEEEtran.bst: No hyphenation pattern has been}%
\typeout{** loaded for the language `#1'. Using the pattern for}%
\typeout{** the default language instead.}%
\else
\language=\csname l@#1\endcsname
\fi
#2}}
\providecommand{\BIBdecl}{\relax}
\BIBdecl

\bibitem{fumanal2021motor}
J.~Fumanal-Idocin, Y.-K. Wang, C.-T. Lin, J.~Fern{\'a}ndez, J.~A. Sanz, and H.~Bustince, ``Motor-imagery-based brain--computer interface using signal derivation and aggregation functions,'' \emph{IEEE Transactions on Cybernetics}, vol.~52, no.~8, pp. 7944--7955, 2021.

\bibitem{li2020sliding}
H.~Li, L.~Bi, and J.~Yi, ``Sliding-mode nonlinear predictive control of brain-controlled mobile robots,'' \emph{IEEE Transactions on Cybernetics}, vol.~52, no.~6, pp. 5419--5431, 2020.

\bibitem{lin2020direct}
C.-T. Lin and T.-T.~N. Do, ``Direct-sense brain--computer interfaces and wearable computers,'' \emph{IEEE Transactions on Systems, Man, and Cybernetics: Systems}, vol.~51, no.~1, pp. 298--312, 2020.

\bibitem{samejima2021brain}
S.~Samejima, A.~Khorasani, V.~Ranganathan, J.~Nakahara, N.~M. Tolley, A.~Boissenin, V.~Shalchyan, M.~R. Daliri, J.~R. Smith, and C.~T. Moritz, ``Brain-computer-spinal interface restores upper limb function after spinal cord injury,'' \emph{IEEE Transactions on Neural Systems and Rehabilitation Engineering}, vol.~29, pp. 1233--1242, 2021.

\bibitem{edelman2024non}
B.~J. Edelman, S.~Zhang, G.~Schalk, P.~Brunner, G.~M{\"u}ller-Putz, C.~Guan, and B.~He, ``Non-invasive brain-computer interfaces: State of the art and trends,'' \emph{IEEE Reviews in Biomedical Engineering}, 2024.

\bibitem{ke2023enhancing}
Y.~Ke, S.~Liu, and D.~Ming, ``Enhancing ssvep identification with less individual calibration data using periodically repeated component analysis,'' \emph{IEEE Transactions on Biomedical Engineering}, 2023.

\bibitem{yin2014dynamically}
E.~Yin, Z.~Zhou, J.~Jiang, Y.~Yu, and D.~Hu, ``A dynamically optimized ssvep brain--computer interface (bci) speller,'' \emph{IEEE transactions on biomedical engineering}, vol.~62, no.~6, pp. 1447--1456, 2014.

\bibitem{guo2022ssvep}
N.~Guo, X.~Wang, D.~Duanmu, X.~Huang, X.~Li, Y.~Fan, H.~Li, Y.~Liu, E.~H.~K. Yeung, M.~K.~T. To \emph{et~al.}, ``Ssvep-based brain computer interface controlled soft robotic glove for post-stroke hand function rehabilitation,'' \emph{IEEE Transactions on Neural Systems and Rehabilitation Engineering}, vol.~30, pp. 1737--1744, 2022.

\bibitem{rivera2022cca}
H.~Rivera-Flor, D.~Gurve, A.~Floriano, D.~Delisle-Rodriguez, R.~Mello, and T.~Bastos-Filho, ``Cca-based compressive sensing for ssvep-based brain-computer interfaces to command a robotic wheelchair,'' \emph{IEEE Transactions on Instrumentation and Measurement}, vol.~71, pp. 1--10, 2022.

\bibitem{lin2006frequency}
Z.~Lin, C.~Zhang, W.~Wu, and X.~Gao, ``Frequency recognition based on canonical correlation analysis for ssvep-based bcis,'' \emph{IEEE transactions on biomedical engineering}, vol.~53, no.~12, pp. 2610--2614, 2006.

\bibitem{chen2015filter}
X.~Chen, Y.~Wang, S.~Gao, T.-P. Jung, and X.~Gao, ``Filter bank canonical correlation analysis for implementing a high-speed ssvep-based brain--computer interface,'' \emph{Journal of neural engineering}, vol.~12, no.~4, p. 046008, 2015.

\bibitem{yuan2015enhancing}
P.~Yuan, X.~Chen, Y.~Wang, X.~Gao, and S.~Gao, ``Enhancing performances of ssvep-based brain--computer interfaces via exploiting inter-subject information,'' \emph{Journal of neural engineering}, vol.~12, no.~4, p. 046006, 2015.

\bibitem{wong2020learning}
C.~M. Wong, F.~Wan, B.~Wang, Z.~Wang, W.~Nan, K.~F. Lao, P.~U. Mak, M.~I. Vai, and A.~Rosa, ``Learning across multi-stimulus enhances target recognition methods in ssvep-based bcis,'' \emph{Journal of neural engineering}, vol.~17, no.~1, p. 016026, 2020.

\bibitem{nakanishi2017enhancing}
M.~Nakanishi, Y.~Wang, X.~Chen, Y.-T. Wang, X.~Gao, and T.-P. Jung, ``Enhancing detection of ssveps for a high-speed brain speller using task-related component analysis,'' \emph{IEEE Transactions on Biomedical Engineering}, vol.~65, no.~1, pp. 104--112, 2017.

\bibitem{obermaier2001information}
B.~Obermaier, C.~Neuper, C.~Guger, and G.~Pfurtscheller, ``Information transfer rate in a five-classes brain-computer interface,'' \emph{IEEE Transactions on neural systems and rehabilitation engineering}, vol.~9, no.~3, pp. 283--288, 2001.

\bibitem{wong2020inter}
C.~M. Wong, Z.~Wang, B.~Wang, K.~F. Lao, A.~Rosa, P.~Xu, T.-P. Jung, C.~P. Chen, and F.~Wan, ``Inter-and intra-subject transfer reduces calibration effort for high-speed ssvep-based bcis,'' \emph{IEEE Transactions on Neural Systems and Rehabilitation Engineering}, vol.~28, no.~10, pp. 2123--2135, 2020.

\bibitem{bian2022small}
R.~Bian, H.~Wu, B.~Liu, and D.~Wu, ``Small data least-squares transformation (sd-lst) for fast calibration of ssvep-based bcis,'' \emph{IEEE Transactions on Neural Systems and Rehabilitation Engineering}, vol.~31, pp. 446--455, 2022.

\bibitem{cao2014objective}
T.~Cao, F.~Wan, C.~M. Wong, J.~N. da~Cruz, and Y.~Hu, ``Objective evaluation of fatigue by eeg spectral analysis in steady-state visual evoked potential-based brain-computer interfaces,'' \emph{Biomedical engineering online}, vol.~13, pp. 1--13, 2014.

\bibitem{wong2021transferring}
C.~M. Wong, Z.~Wang, A.~C. Rosa, C.~P. Chen, T.-P. Jung, Y.~Hu, and F.~Wan, ``Transferring subject-specific knowledge across stimulus frequencies in ssvep-based bcis,'' \emph{IEEE Transactions on Automation Science and Engineering}, vol.~18, no.~2, pp. 552--563, 2021.

\bibitem{wang2022stimulus}
Z.~Wang, C.~M. Wong, A.~Rosa, T.~Qian, T.-P. Jung, and F.~Wan, ``Stimulus-stimulus transfer based on time-frequency-joint representation in ssvep-based bcis,'' \emph{IEEE Transactions on Biomedical Engineering}, vol.~70, no.~2, pp. 603--615, 2022.

\bibitem{huang1998empirical}
N.~E. Huang, Z.~Shen, S.~R. Long, M.~C. Wu, H.~H. Shih, Q.~Zheng, N.-C. Yen, C.~C. Tung, and H.~H. Liu, ``The empirical mode decomposition and the hilbert spectrum for nonlinear and non-stationary time series analysis,'' \emph{Proceedings of the Royal Society of London. Series A: mathematical, physical and engineering sciences}, vol. 454, no. 1971, pp. 903--995, 1998.

\bibitem{zheng2023enhancing}
X.~Zheng, X.~Zhang, G.~Xu, and R.~Zhang, ``Enhancing performance of single-channel ssvep-based visual acuity assessment via mode decomposition,'' \emph{IEEE Transactions on Neural Systems and Rehabilitation Engineering}, 2023.

\bibitem{chen2017new}
Y.-F. Chen, K.~Atal, S.-Q. Xie, and Q.~Liu, ``A new multivariate empirical mode decomposition method for improving the performance of ssvep-based brain--computer interface,'' \emph{Journal of neural engineering}, vol.~14, no.~4, p. 046028, 2017.

\bibitem{chang2022novel}
C.-T. Chang and C.-H. Huang, ``Novel method of multi-frequency flicker to stimulate ssvep and frequency recognition,'' \emph{Biomedical Signal Processing and Control}, vol.~71, p. 103243, 2022.

\bibitem{ouelha2017efficient}
S.~Ouelha, S.~Touati, and B.~Boashash, ``An efficient inverse short-time fourier transform algorithm for improved signal reconstruction by time-frequency synthesis: Optimality and computational issues,'' \emph{Digital Signal Processing}, vol.~65, pp. 81--93, 2017.

\bibitem{lin1996neural}
C.-T. Lin and C.~G. Lee, \emph{Neural fuzzy systems: a neuro-fuzzy synergism to intelligent systems}.\hskip 1em plus 0.5em minus 0.4em\relax Prentice-Hall, Inc., 1996.

\bibitem{10183374}
L.~Ou, Y.-C. Chang, Y.-K. Wang, and C.-T. Lin, ``Fuzzy centered explainable network for reinforcement learning,'' \emph{IEEE Transactions on Fuzzy Systems}, vol.~32, no.~1, pp. 203--213, 2024.

\bibitem{106218}
C.-T. Lin and C.~Lee, ``Neural-network-based fuzzy logic control and decision system,'' \emph{IEEE Transactions on Computers}, vol.~40, no.~12, pp. 1320--1336, 1991.

\bibitem{jiangIFuzzyTLInterpretableFuzzy2024}
\BIBentryALTinterwordspacing
X.~Jiang, B.~Cao, L.~Ou, Y.-C. Chang, T.~Do, and C.-T. Lin. {{iFuzzyTL}}: {{Interpretable Fuzzy Transfer Learning}} for {{SSVEP BCI System}}. [Online]. Available: \url{https://arxiv.org/abs/2410.12267}
\BIBentrySTDinterwordspacing

\bibitem{luFuzzyMachineLearning2024}
J.~Lu, G.~Ma, and G.~Zhang, ``Fuzzy {{Machine Learning}}: {{A Comprehensive Framework}} and {{Systematic Review}},'' \emph{IEEE Transactions on Fuzzy Systems}, vol.~32, no.~7, pp. 3861--3878, 2024.

\bibitem{shihabudheen2018recent}
K.~Shihabudheen and G.~N. Pillai, ``Recent advances in neuro-fuzzy system: A survey,'' \emph{Knowledge-Based Systems}, vol. 152, pp. 136--162, 2018.

\bibitem{zhang2019fusing}
X.~Zhang, G.~Xu, A.~Ravi, W.~Yan, and N.~Jiang, ``Fusing frontal and occipital eeg features to detect “brain switch” by utilizing convolutional neural network,'' \emph{IEEE Access}, vol.~7, pp. 82\,817--82\,825, 2019.

\bibitem{ming2023new}
G.~Ming, H.~Zhong, W.~Pei, X.~Gao, and Y.~Wang, ``A new grid stimulus with subtle flicker perception for user-friendly ssvep-based bcis,'' \emph{Journal of Neural Engineering}, vol.~20, no.~2, p. 026010, 2023.

\bibitem{xie2012steady}
J.~Xie, G.~Xu, J.~Wang, F.~Zhang, and Y.~Zhang, ``Steady-state motion visual evoked potentials produced by oscillating newton's rings: implications for brain-computer interfaces,'' \emph{Plos one}, vol.~7, no.~6, p. e39707, 2012.

\bibitem{waytowich2016optimization}
N.~R. Waytowich, Y.~Yamani, and D.~J. Krusienski, ``Optimization of checkerboard spatial frequencies for steady-state visual evoked potential brain--computer interfaces,'' \emph{IEEE Transactions on Neural Systems and Rehabilitation Engineering}, vol.~25, no.~6, pp. 557--565, 2016.

\bibitem{wang2016benchmark}
Y.~Wang, X.~Chen, X.~Gao, and S.~Gao, ``A benchmark dataset for ssvep-based brain--computer interfaces,'' \emph{IEEE Transactions on Neural Systems and Rehabilitation Engineering}, vol.~25, no.~10, pp. 1746--1752, 2016.

\bibitem{zhou2022empirical}
W.~Zhou, Z.~Feng, Y.~Xu, X.~Wang, and H.~Lv, ``Empirical fourier decomposition: An accurate signal decomposition method for nonlinear and non-stationary time series analysis,'' \emph{Mechanical Systems and Signal Processing}, vol. 163, p. 108155, 2022.

\bibitem{ZHANG2021105572}
\BIBentryALTinterwordspacing
Z.~Zhang, J.~Ding, C.~Zhu, X.~Chen, J.~Wang, L.~Han, X.~Ma, and D.~Xu, ``Bivariate empirical mode decomposition of the spatial variation in the soil organic matter content: A case study from nw china,'' \emph{CATENA}, vol. 206, p. 105572, 2021. [Online]. Available: \url{https://www.sciencedirect.com/science/article/pii/S0341816221004306}
\BIBentrySTDinterwordspacing

\bibitem{jiang2024fuzzybasedapproachpredicthuman}
X.~Jiang, L.~Ou, Y.~Chen, N.~Ao, Y.-C. Chang, T.~Do, and C.-T. Lin, ``A fuzzy logic-based approach to predict human interaction by functional near-infrared spectroscopy,'' \emph{IEEE Transactions on Fuzzy Systems}, pp. 1--15, 2025.

\bibitem{jiang2025iFuzzyAffduo}
X.~Jiang, Y.~Chen, N.~R. Pal, Y.-C. Chang, Y.~Yang, T.~Do, and C.-T. Lin, ``Interpretable dual-filter fuzzy neural networks for affective brain-computer interfaces,'' in \emph{Proceedings of the IEEE International Conference on Fuzzy Systems (FUZZ-IEEE)}.\hskip 1em plus 0.5em minus 0.4em\relax Reims, France: IEEE, 2025, under review.

\bibitem{nakanishi2015comparison}
M.~Nakanishi, Y.~Wang, Y.-T. Wang, and T.-P. Jung, ``A comparison study of canonical correlation analysis based methods for detecting steady-state visual evoked potentials,'' \emph{PloS one}, vol.~10, no.~10, p. e0140703, 2015.

\bibitem{lutovac2001filter}
M.~D. Lutovac, D.~V. To{\v{s}}i{\'c}, and B.~L. Evans, \emph{Filter design for signal processing using MATLAB and Mathematica}.\hskip 1em plus 0.5em minus 0.4em\relax Miroslav Lutovac, 2001.

\bibitem{li2023precise}
H.~Li, G.~Xu, Z.~Li, K.~Zhang, X.~Zheng, C.~Du, C.~Han, J.~Kuang, Y.~Du, and S.~Zhang, ``A precise frequency recognition method of short-time ssvep signals based on signal extension,'' \emph{IEEE Transactions on Neural Systems and Rehabilitation Engineering}, vol.~31, pp. 2486--2496, 2023.

\bibitem{loshchilov2018decoupled}
I.~Loshchilov and F.~Hutter, ``Decoupled weight decay regularization,'' in \emph{International Conference on Learning Representations}, 2019.

\bibitem{kothe2014lab}
C.~Kothe \emph{et~al.}, ``Lab streaming layer (lsl),'' 2014.

\bibitem{postel1980rfc0768}
J.~Postel, ``Rfc0768: User datagram protocol,'' 1980.

\bibitem{cohen2009pearson}
I.~Cohen, Y.~Huang, J.~Chen, J.~Benesty, J.~Benesty, J.~Chen, Y.~Huang, and I.~Cohen, ``Pearson correlation coefficient,'' \emph{Noise reduction in speech processing}, pp. 1--4, 2009.

\end{thebibliography}
\end{document}


\label{SupplementaryMaterial}
\maketitle
\section{Methods and materials}
\subsection{Dataset}
\label{sec:dataset}

\subsubsection{Dataset 1-Benchmark}
The first dataset is Benchmark dataset \cite{wang2016benchmark}, which consists of data collected from 35 subjects. This dataset contains forty frequencies, ranging from 8 Hz to 15.8 Hz, with a 0.2 Hz interval between adjacent frequencies. The phase of the lowest frequency (8 Hz) is set to 0, and it increases by increments of 0.5 $\pi$ as the frequency increases. For each subject, 6 rounds of experiments were conducted. In one round of the experiment, the subject viewed the flicker of each frequency once, resulting in a total of 40 trials per round. Each trial lasts for 6 seconds, comprising an initial cue period of 0.5 seconds, a flicker period of 5 seconds, and a final rest period of 0.5 seconds. Data acquisition of this dataset was implemented with a 64-channel Synamps2 EEG system (Neuroscan, Inc.). The data were then downsampled from 1000 Hz to 250 Hz. In the model training period, we selected ten occipital electrodes including PO3, PO4, PO5, PO6, POZ, O1, O2, OZ, CB1, CB2.

\subsubsection{Dataset 2-12JFPM}
The second dataset is 12JFPM \cite{nakanishi2015comparison}, includes data from 10 subjects. This dataset consists of 12 frequencies ranging from 9.25 Hz to 14.75Hz in increments of 0.5 Hz. The phase of this dataset increase with frequency in increments of 0.5 $\pi$. Experiment of 12JFPM. The experiment in this dataset consisted of 15 rounds, each comprising 12 trials (covering each frequency). The task duration for each trial was 4 seconds. The experimental paradigm was displayed on a 27-inch LCD monitor (ASUS VG278) with a refresh rate of 60 Hz and a resolution of 1280×800 pixels. Data from eight occipital channels were collected using the BioSemi ActiveTwo EEG systems (BioSemi B.V.) at a sampling rate of 2048 Hz, which was subsequently downsampled to 256 Hz. 

\subsection{The Baseline Methods}
\label{sec:baseline}

\subsubsection{TFJR}
For TFJR \cite{wang2022stimulus}, a DPMAFD-based adaptive decomposition technique is proposed for the simultaneous adaptive decomposition of SSVEP signals from various stimuli. Initially, the DPMAFD is utilized to decompose the source domain data into various components. Following decomposition, the coefficients corresponding to common components across the source domain are identified and preserved. Redundant noise components are identified and eliminated from the analysis. In the final step, the inverse DPMAFD is employed to reconstruct the target domain SSVEP signals. TFJR can identify and select major common periodic components across different stimuli, which are crucial for enhancing the transfer learning process.

\subsubsection{tlCCA}
For tlCCA \cite{wong2021transferring}, initially, it applies a linear decomposition method based on the superposition theory to separate the SSVEP signal into impulse response and  periodic impulse. Then the impulse response from the source domain is combined with artificially created periodic impulses of the target domain. Next, the combined components (impulse response and fabricated periodic impulses) are then merged with convolution to reconstruct synthetic signals that are tailored for the target domain. Finally, spatial filters of classifier are trained with the source-domain data and the reconstructed data.

\subsubsection{EMD-ECCA}
ECCA is an SSVEP classification method that improves upon the classical CCA approach \cite{wong2020inter}. It averages existing data to generate subject-specific SSVEP templates. These templates, along with standard sine and cosine signals, are used as reference signals in CCA calculations with new SSVEP data, facilitating more accurate classification.
Similarly, in this study, the dataset consisting of target-domain signals reconstructed using EMD, along with source-domain signals, is employed as the input for ECCA to facilitate model training.

\subsubsection{FBCCA}
Initially, FBCCA leverages a bank of band-pass filters to SSVEP signals to decompose them into multiple sub-bands \cite{chen2015filter}. Followed by filtering, CCA is performed on each sub-band separately. The highest correlation coefficients from each sub-band are then aggregated to determine the dominant frequency of the SSVEP response. Finally, a decision rule based on these coefficients classifies the target stimulus frequency.

\bibliographystyle{IEEEtran}
\bibliography{ref, ref_Beining}